\documentclass[a4paper,11pt]{article}
\usepackage{jheppub} 
\usepackage{lineno}
\usepackage{shuffle}
\usepackage{float}
\usepackage{subcaption}

\title{\boldmath The $gg\rightarrow HH$ amplitude induced by bottom quarks at two-loop level: planar master integrals}

\author[a,b]{Zhenghong Hu,}
\author[a,b]{Tao Liu,}
\author[c,b,d]{Jin Min Yang}
 
\affiliation[a]{Institute of High Energy Physics, Chinese Academy of Sciences, Beijing 100049, P. R. China}
\affiliation[b]{School of Physical Sciences, University of Chinese Academy of Sciences, Beijing 100049, P. R. China}
\affiliation[c]{Institute of Theoretical Physics, Chinese Academy of Sciences, Beijing 100190, P. R. China }
\affiliation[d]{Center for Theoretical Physics, Henan Normal University, Xinxiang 453007,  P. R. China}

\emailAdd{huzh@ihep.ac.cn} 
\emailAdd{liutao86@ihep.ac.cn}
\emailAdd{jmyang@itp.ac.cn}

\abstract{We consider the two-loop amplitude of $gg\rightarrow HH$ mediated by bottom quarks, which provides a correction of percent level at leading order in the low invariant mass region. In order to compute the corresponding master integrals, we perform an expansion for the bottom quark mass, using the method of differential equations and fixing the boundary constants by a numerical approach. The results of 177 planar master integrals which are expressed in terms of multiple polylogarithms are provided. 
}

\begin{document}
\maketitle
\flushbottom
    
\section{Introduction}
Higgs pair production via gluon fusion is the most promising process to access the trilinear Higgs coupling at the Large Hadron Collider (LHC).   
Thus it is important to provide precise theoretical predictions and there are many studies on QCD and electroweak (EW) corrections to $gg\rightarrow HH$.   
The leading order (LO) results were computed as early as last century~\cite{Glover:1987nx, Plehn:1996wb}. 
At next-to-leading order (NLO) both QCD and EW corrections require the computation of two-loop four-point Feynman integrals with massive propagators which are challenging technically. Comprehensive NLO results that consider the effect of all mass scales have only been known with the two-loop virtual amplitudes calculated by numerical methods. The QCD corrections are evaluated in Refs.~\cite{Borowka:2016ehy,Borowka:2016ypz,Baglio:2018lrj,Davies:2019dfy}, and EW corrections in Refs.~\cite{Bi:2023bnq,Heinrich:2024dnz}. To avoid the well-known drawbacks of numerical evaluations and reduce the scales involved during the calculation, there are a number of attempts in the direction of analytical expansion. 
Analytical QCD corrections at NLO have been obtained through performing Taylor expansions or asymptotic expansions in different kinematic limits\footnote{In Refs.~\cite{Xu:2018eos,Wang:2020nnr} an expansion has been performed for small Higgs boson mass and then the elliptic Feynman integrals are evaluated in numerical ways.}, which include large top quark mass limit~\cite{Dawson:1998py,Grigo:2013rya,Degrassi:2016vss,Muhlleitner:2022ijf}, top quark threshold limit~\cite{Grober:2017uho}, the high-energy limit~\cite{Davies:2018ood,Davies:2018qvx,Davies:2019dfy}, small transverse momentum limit~\cite{Bonciani:2018omm} and forward-scattering limit~\cite{Davies:2023vmj}. 
It is shown in Ref.~\cite{Bellafronte:2022jmo} that combining results of different limits could even provide valid amplitudes that nearly cover the entire phase space. 
Similar to NLO QCD contributions, EW corrections at NLO have been calculated analytically in Refs.~\cite{Davies:2022ram,Davies:2023npk,Davies:2025wke,Bonetti:2025vfd}.
There are also studies performed by different groups beyond NLO, e.g., the NNLO QCD corrections to the amplitude are considered in Refs.~\cite{Davies:2019djw,Davies:2023obx,Grazzini:2018bsd, Davies:2024znp, Davies:2025ghl}. In the large top quark mass limit $\text{N}^3\text{LO}$ QCD corrections are also available~\cite{Chen:2019lzz,Chen:2019fhs}.

What we focus on in this paper is the amplitude induced by bottom quarks whose mass is small compared to the mass of Higgs boson\footnote{Obviously our result applies to other light quarks as well.}. Note that at one-loop level the bottom quark contribution to single Higgs production is suppressed by $m_b^2/m_h^2$ but enhanced by $\ln^2 (m_b^2/m_h^2)$ relative to the top quark contribution, which leads to a percent level correction to the total cross section~\cite{Liu:2017vkm,Liu:2018czl,Anastasiou:2020vkr}. It is natural to expect this kind of logarithms and corrections of the same order of magnitude from bottom quarks for Higgs pair production, at least in certain regions of phase space.
In Table~\ref{tab:tabI} we show cross sections under different cuts on the invariant mass of Higgs pair at the 13 TeV LHC. We see that the interferences between bottom and top quarks could provide percent level corrections in the region of low invariant mass.
As for the NLO QCD corrections, numerical calculations in principle should not be a big problem with the help of the package {\tt AMFlow}~\cite{Liu:2017jxz,Liu:2022mfb,Liu:2022chg}, since it has been successfully used for calculating the NLO electroweak corrections~\cite{Bi:2023bnq}. We aim to get an analytical expression of the NLO amplitude, since it may be helpful to understand the underlying logarithmic structures just like the analysis on the two-loop $gg\rightarrow hg$ amplitude in Ref.~\cite{Liu:2024tkc,Melnikov:2016qoc,Lindert:2017pky,Bonciani:2022jmb}, which is also mediated by bottom quarks. There is also a logarithmic analysis on the top quark contribution to Higgs pair production in the high energy limit, where the analytical results prove to be useful~\cite{Jaskiewicz:2024xkd}.   
Another motivation is that the master integrals we get from the calculation could also be used for other physical processes induced by light quarks, such as $gg\rightarrow ZZ$~\cite{Agarwal:2024pod}. As the first step towards this goal, in this paper we consider the part of the amplitude which can be expressed in terms of planar master integrals.

\begin{table}[]
    \centering
    \begin{tabular}{c|c|c|c}
        \hline
        $M^{max}_{HH}$  & $\sigma_{t}$(fb) & $\sigma_{t,b}$(fb) & $\frac{\sigma_{t,b}-\sigma_{t}}{\sigma_{t}}$ \\
         \hline
        300 & 0.229(3) & 0.242(6) & 5.7\% \\
        \hline
        350 & 1.38(0) & 1.41(3) & 2.2\%   \\
        \hline
        400 & 4.33(2) & 4.37(7)  & 0.92\% \\
        \hline
    \end{tabular}
    \caption{Cross sections of Higgs pair production at the LO with $\sqrt{s}=13\,\text{TeV}$ and different cuts on the invariant mass $M^{max}_{HH}$. These results are obtained with the help of {\tt MadGraph5\_aMC@NLO}~\cite{Alwall:2014hca}. The factorization and renormalization scales are set to be the same as the dynamical partonic center-of-mass energy and we use NNPDF3.1~\cite{NNPDF:2017mvq} with $\rm NNPDF31\underline{~}nlo\underline{~}as\underline{~}0118$ for parton distribution functions.}
    \label{tab:tabI}
\end{table}

The rest of this paper is organized as follows. In the next section we introduce our notations, present technical details on the amplitudes and the reduction of Feynman integrals. In Section~\ref{Sec:calc}, we explain the differential equation method used for calculating the master integrals and discuss how to get the boundary constants through numerical ways.       
The conclusion is made in Section~\ref{Sec:con}. 
In the Appendix, we provide the details of the master integrals and give a brief introduction to multiple polylogarithms.

\section{Notations and master integrals}

We consider the process mediated by a bottom quark loop
\begin{equation}
    g(p_1)g(p_2) \rightarrow H(p_3) H(p_4), 
\end{equation}
where all momenta are incoming. The masses of bottom quark and Higgs boson are denoted by $m_b$ and $m_H$ respectively. The Mandelstam variables are defined as
\begin{equation}
    s = (p_1+p_2)^2\,,\quad t=(p_1+p_3)^2\,, \quad u=(p_2+p_3)^2\,,
\end{equation}
with the relations
\begin{equation}
    p_1^2=p_2^2=0\,,\quad p_3^2=p_4^2=m_H^2\,, \quad s+t+u=2m_H^2\,.
\end{equation}
The amplitude will depend on four independent variables $s,\,u,\,m_b^2$ and $m_H^2$. 

Next, we define two dimensionless variables $x$ and $z$ following Ref.~\cite{Gehrmann:2014bfa}, in which master integrals with  massless internal quarks are calculated, and an additional one $\kappa$  
\begin{equation}
    s=m_H^2 \, \frac{(1+x)^2}{x}\,, \quad u=-m_H^2 \, z\,, \quad m_b^2=m_H^2 \, \kappa\,.
    \label{eq:def var}
\end{equation}
Considering the amplitude in the physical region, we have that 
\begin{equation}
   1> x > 0 \,, \quad z > 0\,, \quad \kappa > 0 \,, \quad m_H^2>0\,.
\end{equation}

\begin{figure}
    \centering
    \includegraphics[width=0.5\linewidth]{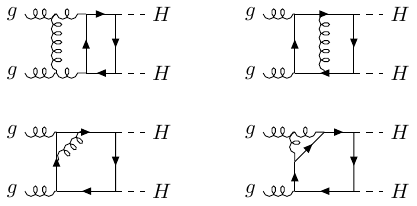}
    \caption{Sample Feynman diagrams for the two-loop QCD corrections of $gg \rightarrow H H$. Solid lines with arrow represent intermediate quarks, while curly and dashed lines represent gluons and Higgs respectively.}
    \label{fig:feynman}
\end{figure}

There are two linearly independent Lorentz structures for the amplitude, and we can write two corresponding form factors,
\begin{eqnarray}
  {\cal M}^{\mu\nu} = \left( {\cal M}_1 A_1^{\mu\nu} + {\cal M}_2 A_2^{\mu\nu} \right) \,,
\end{eqnarray}
with
\begin{eqnarray}
  A_1^{\mu\nu} &=& g^{\mu\nu} - \frac{1}{p_{12}}p_1^\nu p_2^\mu\,,\nonumber\\
  A_2^{\mu\nu} &=& g^{\mu\nu} 
  + \frac{ p_{33} }{ p_T^2 p_{12} } p_1^\nu p_2^\mu    
  - \frac{ 2p_{23} }{ p_T^2 p_{12} } p_1^\nu p_3^\mu    
  - \frac{ 2p_{13} }{ p_T^2 p_{12} } p_3^\nu p_2^\mu    
  + \frac{ 2 }{ p_T^2 } p_3^\nu p_3^\mu    
  \,,
\end{eqnarray}
where $p_{ij} = p_i\cdot p_j$ and $p_T^2=\frac{u \, t -m_H^4}{s}$. With the help of the projectors $P_{i,\mu \nu}$ one can obtain the form factors by ${\cal M}_i = P_{i,\mu\nu} {\cal M}^{\mu\nu}$. The explicit expressions of these projectors can be found in Ref.~\cite{Davies:2018ood} and they are not listed here for simplicity.
Furthermore, we can decompose the two form factors into ``triangle" and ``box" parts, 
\begin{eqnarray}
  {\cal M}_1 &=& X_0 \, s \, \left(\frac{3 m_H^2}{s-m_H^2} F_{\rm tri} + F_{\rm box1}\right)
                 \,,\nonumber\\
  {\cal M}_2 &=& X_0 \, s \, F_{\rm box2}
                 \,,
\end{eqnarray}
with
\begin{eqnarray}
  X_0 &=& \frac{G_F}{\sqrt{2}} \frac{\alpha_s(\mu)}{2\pi} T \,.
\end{eqnarray}
Here $T=1/2$, $\mu$ is the renormalization scale and $G_F$ is the Fermi constant. Note that there are also ``double-triangle''contribution coming from Feynman diagrams with two closed quark loops at NLO. This kind of form factors is easy to get since only one-loop integrals need to be evaluated. Feynman diagrams involving Higgs self-couplings only contribute to the triangle part $F_{tri}$ and its NLO corrections has been computed in Refs.~\cite{Harlander:2005rq,Anastasiou:2006hc,Aglietti:2006tp}. In this paper, we will focus on the two-loop QCD corrections to the ``box" part, typical Feynman diagrams of which are shown in Fig.~\ref{fig:feynman}. 

To compute the form factors, {\tt QGRAF}~\cite{NOGUEIRA1993279} is used for the generation of two-loop Feynman diagrams of box type. We use {\tt FORM}~\cite{Vermaseren:2000nd} to apply aforementioned projectors and perform Dirac and color algebra on each diagram. Then we are left with linear combinations of scalar Feynman integrals which could be categorized into 22 integral families. For planar diagrams there are 10 integral families to consider and 7 other families can be generated through permutation of the external momenta $p_1 \leftrightarrow p_2$ or $p_3 \leftrightarrow p_4$. 
Using dimensional regularization $d=4-2\epsilon$, each integral is defined as
\begin{equation}
 \text{I}^{\text{PL}}(a_1,a_2,...,a_8,a_9) = \left(\frac{S_\epsilon}{16 \pi^2}\right)^{-2}\, (m_H^2)^{2 \epsilon }
\int \frac{d^d k_1}{(2 \pi)^d}\frac{d^d k_2}{(2 \pi)^d}\,
  \frac{1}{D_1^{a_1}\,D_2^{a_2}...D_8^{a_8}\,D_9^{a_9}}\,
\label{eq:def int}
\end{equation}
with
\begin{equation}
  S_\epsilon =  (4 \pi)^\epsilon \, \frac{\Gamma(1+ \epsilon) \, 
\Gamma^2(1-\epsilon)}{\Gamma(1-2\epsilon)}\,,
\end{equation}
where ``PL" represents an integral family with propagators $D_i$. Here $a_{1-7}$ are non-negative integers and $a_{8,9}$ are non-positive ones. Detailed information on the integral families and their corresponding propagators can be found in Appendix~\ref{appendixA}. In our calculations one-loop integral families are considered first, which serve as a test of the differential equations method we use and they are also provided in the Appendix.

It is well known that scalar integrals can be reduced to master integrals (MIs) by applying integration by parts identities (IBPs)~\cite{TKACHOV198165,Chetyrkin:1981qh} and this procedure has already been implemented in many publicly available programs, such as $\tt FIRE$~\cite{Smirnov:2019qkx,Smirnov:2023yhb}, $\tt LiteRed$~\cite{Lee:2013mka} and $\tt Kira$~\cite{Maierhofer:2017gsa,Klappert:2020nbg}. Here $\tt Kira$, which is based on Laporta’s algorithm~\cite{Laporta:2000dsw} and employs the method of finite fields with the help of $\tt FireFly$~\cite{Klappert:2019emp,Klappert:2020aqs}, is used for our reduction. For simplification we set $m_H=1$ and reconstruct the mass dimension
when necessary throughout the whole calculation, so there are three free variables $s\,,u$ and $m_b^2$ existing during the reduction. These variables will be transformed to $x\,,z$ and $\kappa$ when constructing differential equations with the help of the reduction results. Finally, we get 177 MIs for two-loop planar integral families and 11 MIs for one-loop families, which are listed in Appendix~\ref{appendixA}.
  
\section{Calculation of master integrals}
\label{Sec:calc}
The differential equations method~\cite{Kotikov:1990kg,Kotikov:1991pm} has proven to be one of the most powerful tools to compute Feynman integrals or MIs, especially after the breakthrough was made in Ref.~\cite{Henn:2013pwa} which proposed the idea of canonical form of differential equations.  
However, as mentioned for example in \cite{Davies:2018ood} it is not easy to find the canonical basis of all sectors for their master integrals. We believe that the same problem would appear in our case since our integral families are the same as theirs except for different treatment of the external Higgs bosons.  
Thus we do not insist on finding the canonical basis, but choose to solve differential equations using the ansatz guided by asymptotic expansion \cite{Beneke:1997zp,Smirnov:2002pj} which has been successfully used in Refs.~\cite{Melnikov:2016qoc,Davies:2018ood,Davies:2018qvx}. In this ansatz the small bottom quark mass will be expanded and separated from kinematical scales $s,u$. It is expected that now   
all the information of $s,u$ is contained in a class of well-understood special functions, namely multiple polylogarithms (MPLs)~\cite{goncharov2011multiplepolylogarithmscyclotomymodular,goncharov2001multiplepolylogarithmsmixedtate}. Solutions of differential equations with respect to the kinematical scales would determine the exact form of MPLs for the MIs, after we get the relations of the unknown coefficients in the ansatz from differential equations with respect to $m_b$. At the last step we will get the analytical results once the boundary conditions are fixed.   
In the following subsections we will explain all the details on what we do. 
 
\subsection{Differential equations}
The idea of differential equations is that when we take a derivative of a master integral with respect to a kinematic variable or internal quark mass, we obtain results expressed in terms of integrals of its own family. By applying IBPs, we can transform them into a linear combination of the MIs. This allows us to construct a system of differential equations which can be solved with unknown boundary conditions. 

As mentioned in the last section, we keep $s\,,u\,,m_H^2$ and $m_b^2$ as independent variables, and we will also construct  partial differential equations with respect to them. The derivatives with respect to $s,u$ and $m_H^2$ can be expressed as the derivatives with respect to the external momenta, which yield integrals of another set of $\{a_i\}$ in Eq.~(\ref{eq:def int}). Specifically, we have
\begin{eqnarray}
   \partial_{s} &=& \frac{l_3}{2 s l_1 l_2} p_1 \cdot \partial_{p_1} + \frac{\left(u- m_H^2 \right) l_4}{2s l_1 l_2} p_2 \cdot \partial_{p_2} - \frac{\left(u- m_H^2 \right)^2}{2s l_2} p_3 \cdot \partial_{p_3}\nonumber\\
   &-& \frac{m_H^2 \left(u-m_H^2\right)^2}{s l_1 l_2} p_1 \cdot \partial_{p_3}, \nonumber \\
   \partial_{u} &=& \frac{s l_5}{2 l_1 l_2} p_1 \cdot \partial_{p_1} - \frac{s l_5}{2 l_1 l_2}p_2 \cdot \partial_{p_2} + \frac{l_6}{2 l_2}p_3 \cdot \partial_{p_3} + \frac{m_H^2 l_6}{l_1 l_2} p_1 \cdot \partial_{p_3},\nonumber\\
   \partial_{m_H^2} &=& \frac{-m_H^2 s}{l_1 l_2} p_1\cdot \partial_{p_1} +\frac{m_H^2 s}{l_1 l_2} p_2 \cdot \partial_{p_2} + \frac{m_H^2-u}{l_2} p_3 \cdot \partial_{p_3}+\frac{l_7}{l_1 l_2} p_1 \cdot \partial_{p_3},
\end{eqnarray}
where
\begin{eqnarray}
    l_1 & = &  s+u-m_H^2, \nonumber \\
    l_2 & = & m_H^4-2 m_H^2 u+u (s+u),\nonumber\\
    l_3 & = & 3 m_H^4 (s+u)-m_H^6-3 m_H^2 u (2 s+u)+u (s+u) (2 s+u), \nonumber \\
    l_4 & = & m_H^4+m_H^2 (s-2 u)+u (s+u) , \nonumber\\
    l_5 &= & 3 m_H^2-s-u, \nonumber \\
    l_6 & = & s+2 u-2 m_H^2,\nonumber\\
    l_7 & = & m_H^4-u (s+u).
\end{eqnarray}
with $p_i \cdot \partial_{p_j} = p_i^\mu \frac{\partial}{\partial p_j^\mu}$. The derivative with respect to $m_b^2$ is simple, so we will not explain it further. Next, according to Eq.~(\ref{eq:def var}) we can construct differential equations with respect to dimensionless variables $x,z$ and $\kappa$
\begin{equation}
\partial_{x}=\frac{x^2-1}{x^2} m_H^2 \, \partial_{s} \,, \quad \partial_{z}= - m_H^2 \, \partial_{u} \,, \quad \partial_{\kappa}= m_H^2 \, \partial_{m_b^2}\,.
\label{eq:partial}
\end{equation}
Here the dependence on $m_H^2$ is shown explicitly to avoid misunderstanding. 
Note that now there is only one dimensional variable $m_H$ which determines the canonical mass dimension of the MIs, and thus we do not need the differential equations with respect to it.
Finally the useful differential equations take the form
\begin{equation}
\partial_k \text{I}_i (x,z,\kappa,\epsilon)= A^k_{i j}(x,z,\kappa,\epsilon) \, \text{I}_j(x,z,\kappa,\epsilon)\,, \quad k \in \{ x,\, z,\, \kappa \}\,,
\label{eq:DE}
\end{equation}
where $A^k_{i j}$ are rational functions of $x,z,\kappa$ and dimensional regulator $\epsilon$.

It is usually difficult to solve differential equations like  Eq.~(\ref{eq:DE}) with three variables $x,z$ and $\kappa$ directly. Here we use the following ansatz for each master integral which is suitable to describe the solution in the limit of $m_b^2 \rightarrow 0$ or $\kappa \rightarrow 0$, 
\begin{equation}
 \text{I}_i(x,z,\kappa,\epsilon) = \sum_{n=n_{min}}^{n_{max}} \sum_{j=0}^{2} \sum_{k=0}^{2} c^i_{n,j,k}(x,z,\epsilon) \kappa^{n-j\epsilon}  \log^k \kappa\,. 
 \label{eq:asyexp}
\end{equation}
$n_{min}$ depends on the power of the propagators of the master integral and 
the strongest singularity for $\kappa$ we encountered is $\kappa^{-1}$.\footnote{MIs of the non-planar families may contain terms proportional to odd powers of quark mass after expansion and we leave them for future study.} As for $n_{max}$ which in principle can goes to infinity, currently we set it to be one. Note that when terms proportional to $\kappa^0$ and $\kappa^1$ have been solved, high order terms could be immediately obtained through differential equations with respect to $\kappa$. About the coefficients $c^i_{n,j,k}$, it is interesting to see that $c^i_{0,0,0}$ corresponds to the ``all-hard" region in the language of expansion by regions, which is exactly the same integral with the quark mass goes to zero. The massless limit of these integrals has been calculated in Ref.~\cite{Gehrmann:2014bfa} and later they will provide a strong crosscheck for our calculation. 

Next, the differential equations with respect to $\kappa$ will be considered at the first step. 
Substituting Eq.~(\ref{eq:asyexp}) into Eq.~(\ref{eq:DE}) gives
\small
\begin{eqnarray}
    \partial_\kappa \text{I}_i (x,z,\kappa,\epsilon) &=& \sum_{n=n_{min}}^{n_{max}} \sum_{j=0}^{2} \sum_{k=0}^{2} \left \{ (n-j\epsilon) \, c^i_{n,j,k}(x,z,\epsilon) \kappa^{n-j\epsilon -1}  \log^k \kappa \nonumber \right. \\
    && \left. +k \, c^i_{n,j,k}(x,z,\epsilon) \kappa^{n-j\epsilon -1}  \log^{k-1} \kappa \right\} \,, \nonumber \\
    A^\kappa_{i l}(x,z,\kappa,\epsilon) \, \text{I}_l(x,z,\kappa,\epsilon) &=&  \sum_{n=n_{min}}^{n_{max}} \sum_{j=0}^{2} \sum_{k=0}^{2} {\sum_{m}} A^{\kappa,(m)}_{i l}(x,z,\epsilon) \,  c^l_{n,j,k}(x,z,\epsilon) \kappa^{n-j\epsilon + m}  \log^k \kappa, \label{eq:kappa}
\end{eqnarray}
\normalsize 
where $A^{\kappa,(m)}_{i l}$ is the coefficient of $\kappa^m$ in the expansion of $A^\kappa_{i l}$ with respect to $\kappa$. Under the condition that coefficients of $\kappa^{n-j\epsilon}\log^k \kappa$ in the above equations are equal independently, we obtain a system of linear equations for $ c^i_{n,j,k}$.
After solving this linear system, we will be left with some undetermined coefficients. In general, if a sector has $N$ MIs, there will be $N$ undetermined coefficients. 
To determine these coefficients the differential equations with respect to $x$ and $z$ have to be considered and we again demand that different $\kappa^{n-j\epsilon}\log^k \kappa$ terms are independent. 

Before going on, the expansions in $\epsilon$ for $c^i_{n,j,k}$ are necessary for calculating physical scattering amplitudes and we get 
\begin{equation}
    c^i_{n,j,k} (x,z,\epsilon) = \sum_{r=r_{min}}^{r_{max}} \epsilon^r c^{i,(r)}_{n,j,k}(x,z)\label{eq:epsilon} \,,
\end{equation}
where $r_{max}$ takes different values for different master integrals. 
The reason is that, although we require all MIs should be expanded at least to $O(\epsilon)$, higher order terms for some MIs are indispensable since there are $\epsilon$ poles before them during reduction. For this kind of MIs, $r_{max}$ are fixed case by case. 
Now each integral has the form
\begin{equation}
 \text{I}_i(x,z,\kappa,\epsilon) = \sum_{r=r_{min}}^{r_{max}} \sum_{n=n_{min}}^{n_{max}} \sum_{j=0}^{2} \sum_{k=0}^{2} c^{i,(r)}_{n,j,k}(x,z) \epsilon^r \kappa^{n-j\epsilon}  \log^k \kappa\,.
 \label{eq:full exp}
\end{equation}
According to the above discussion, substituting Eq.~(\ref{eq:full exp}) into Eq.~(\ref{eq:DE}) will provide differential equations with respect to $x$ and $z$ for the free coefficients $c^{i,(r)}_{n,j,k}$.

Now we can solve these undetermined coefficients sector by sector. If $j$ of all $c^{i}_{n,j,k}$ are different in a sector, we always obtain a first-order differential equation for $c^{i,(r)}_{n,j,k}$, which can be solved order by order in $\epsilon$ directly. If several undetermined coefficients have the same $j$, we may encounter a system of coupled differential equations.  Below we will give a simple example to illustrate how to solve this kind of equations. Suppose that in a sector we have two coefficients to solve and their differential equations with respect to $x$ take the following form
\begin{eqnarray}
    \frac{\partial}{\partial x} c^{i_1,(r_1)}_{n_1,j,k_1} (x,z) & = & a_{11}(x,z) \, c^{i_1,(r_1)}_{n_1,j,k_1}(x,z) + a_{12}(x,z) \, c^{i_2,(r_2)}_{n_2,j,k_2} (x,z) + b_1(x,z) \,,  \nonumber\\
    \frac{\partial}{\partial x} c^{i_2,(r_2)}_{n_2,j,k_2} (x,z) & = & a_{21}(x,z) \, c^{i_1,(r_1)}_{n_1,j,k_1}(x,z) + a_{22}(x,z) \, c^{i_2,(r_2)}_{n_2,j,k_2} (x,z) + b_2(x,z) \,,
\end{eqnarray}
where $a_{ij}(x,z)$ and $b_i(x,z)$ are known functions of $x$ and $z$. The general way we use is taking a derivative of one of the equations with respect to $x$ and then transforming it into a second-order differential equation,  
\begin{equation}
    \frac{\partial^2}{\partial x^2} c^{i_1,(r_1)}_{n_1,j,k_1} (x,z) = c_1(x,z) \frac{\partial}{\partial x} c^{i_1,(r_1)}_{n_1,j,k_1}(x,z) + c_2(x,z) \, c^{i_1,(r_1)}_{n_1,j,k_1}(x,z) + f(x,z) 
    \label{eq:2nd De}
\end{equation}
with
\begin{eqnarray}
    c_1 & = & a_{11} + a_{22} + \frac{1}{a_{12}} \, \frac{\partial a_{12}}{\partial x} \,,  \nonumber\\
    c_2 & = & a_{12} \, a_{21} - a_{11} \, a_{22} + \frac{\partial a_{11}}{\partial x} -  \frac{a_{11}}{a_{12}} \, \frac{\partial a_{12}}{\partial x} \,,  \nonumber \\
    f & = & a_{12} \, b_{2} -a_{22} \, b_{1} \frac{\partial b_{1}}{\partial x} - \frac{b_{1}}{a_{12}} \, \frac{\partial a_{12}}{\partial x} \,.
\end{eqnarray}
If $c_2$ equals zero, it is just a first-order differential equation for $\frac{\partial}{\partial x} c^{i_1,(r_1)}_{n_1,j,k_1}$. For non-vanishing $c_2$, we could replace $c^{i_1,(r_1)}_{n_1,j,k_1}(x,z)$ with $g(x) \phi (x,z)$ and require  
\begin{equation}
    g''(x) - c_1(x,z) \, g'(x) -c_2(x,z) \, g(x) =0 \,. 
\end{equation}
From Eq.~(\ref{eq:2nd De}) it is easy to find that
\begin{equation}
    \phi''(x,z) + \left(2 g'(x,z)-c_1(x,z) g(x,z)\right)\phi'(x,z)-f(x,z)=0\,.
\end{equation}
Obviously, $g(x)\phi(x,z)$ is much simpler to solve than the original $c^{i_1,(r_1)}_{n_1,j,k_1}$. Another tricky method which may work is constructing a new coefficient and utilizing the relations between the original coefficients. If the result of partial derivative operation on this coefficient does not depend on $c^{i_1,(r_1)}_{n_1,j,k_1}$ or $c^{i_2,(r_2)}_{n_2,j,k_2}$ explicitly, we again obtain a first-order equation.  

In actual calculations we found that it is rather difficult to directly solve and further simplify the solutions of the linear system derived from the differential equations with respect to $\kappa$ due to the complexity of matrix $A^{\kappa,(m)}_{i l}$ in Eq.~(\ref{eq:kappa}). And we use $\tt{FiniteFlow}$~\cite{Peraro:2019svx} to reconstruct the solutions from numerical evaluations over finite fields. After solving the differential equations with respect to $x$ and $z$, we finally obtain the MIs expressed in terms of MPLs with unknown boundary conditions, which will be discussed in the next subsection. The definition and useful properties of MPLs can be found in Appendix~\ref{appendixB}. $\tt{Mathematica}$ package $\tt{PolyLogTools}$~\cite{Duhr:2019tlz} have been used to perform integration, differentiation, and simplification of MPLs. In the next subsection, we will explain how to get the analytical form of boundary conditions by numerical methods.

\subsection{Boundary conditions}
\label{sec:Bc}

According to the definition of MPLs in Eq.~(\ref{eq:def mpl}), boundary conditions are essentially the values of the integrals at the point $x,\,z=0$. Assuming that there is a list of real numbers $\left\{x_1,x_2,...,x_n\right\}$, PSLQ algorithm can find integers $a_{1,...,n}$ with not all $a_i=0$ satisfy the relation
\begin{equation}
    a_1 x_1+a_2 x_2+...+a_n x_n=0\,. 
\end{equation} 
Therefore, we can obtain the analytical form of the boundary using the PSLQ algorithm\footnote{Common numbers(up to weight six) that appear in the calculation of Feynman integrals, such as $\ln2,\pi,\zeta(n),\text{Li}_n(1/2)$ and also the products of them, have been used in our case for planar diagrams.} if we know their numerical values precisely enough. Note that we cannot directly evaluate the master integrals at the boundary, since $x=0$ corresponds to $s=\infty$ which is problematic for numerical evaluations in many public codes. So the best choice is to subtract the contribution of MPLs from the full result at a physical point. In this paper $\tt{AMFlow}$ are used to calculate master integrals to a high precision and for MPLs we choose the C++ library $\tt{GiNaC}$~\cite{bauer2001introductionginacframeworksymbolic}. 
To conveniently identify the branch of MPLs, we require
\begin{equation}
    s \rightarrow s + i 0^+ \,,\quad u \rightarrow u +  i 0^+ \,,\quad m_b^2 \rightarrow m_b^2 - i 0^+ \,,\quad m_H^2 \rightarrow m_H^2 + i 0^+,
\end{equation}
where $0^+$ is an infinitesimal imaginary part.

Another point we have to emphasize is that usually what we get from the subtraction is correct only to a certain degree of accuracy.  
$\tt{AMFlow}$ compute the original MI which equals the right side of Eq.~(\ref{eq:asyexp}) with $n_{max} = \infty$, while MPLs are obtained at fixed order of $\kappa$. 
In order to improve the accuracy we employ two methods. The first one is making $\kappa$ as small as possible and the other is to increase the expansion order of $\kappa$. In real calculations we prioritize the first method by setting $\kappa=10^{-25}$ and demand that stable results within a precision of around 30 digits should be obtained by the PSLQ algorithm. Here we believe that the result is stable if it does depends on the digits of the numerical numbers which are used as an input for this algorithm.
When stable results cannot be obtained in this way, we will use the second method to further enhance the precision until the goal is achieved. We actually encountered this kind of problems during the calculation of the coefficients of high orders in $\epsilon$. 

As mentioned before, the coefficients of $\kappa^0 \log^0 \kappa$ can be directly computed with $\kappa=0$ based on the argument of expansion by regions. For this kind of coefficients or integrals the results we obtained were compared with those in Ref.~\cite{Gehrmann:2014bfa} and we do not find any inconsistencies. This fact gives us enough confidence for our whole calculation. Next an concrete example will be give below to show what we do in real calculations.    

Considering the integral family PL5, which contains 60 MIs.  Solving the differential equations with respect to $\kappa$ results in 60 undetermined coefficients. Suppose that we have derived 56 MIs, there are still four MIs in the top sector and thus four coefficients left.  These coefficients are
\begin{equation}
    c^{57}_{0,-1,0}(x,z) \,, \quad c^{57}_{0,0,0}(x,z) \,, \quad c^{58}_{0,0,0}(x,z) \,, \quad c^{60}_{0,0,0}(x,z) \,.
\end{equation}
From the numerical results in the massless case, we immediately know the values of $r_{min}$ in Eq.~(\ref{eq:epsilon}) for the last three coefficients. $r_{min}$ of $c^{57}_{0,-1,0}$ is just set to be -4 according to the general principle. Then we get   
\begin{eqnarray}
    c^{57}_{0,-1,0}(x,z) & = \sum\limits_{i=-4}^0 c_0^{(i)}(x,z) \epsilon^i + O(\epsilon) \,, \quad 
    c^{57}_{0,0,0}(x,z) & = \sum\limits_{i=-4}^0 c_1^{(i)}(x,z) \epsilon^i + O(\epsilon) \,,  \nonumber\\
    c^{58}_{0,0,0}(x,z) & = \sum\limits_{i=-2}^0 c_2^{(i)}(x,z) \epsilon^i + O(\epsilon) \,, \quad 
    c^{60}_{0,0,0}(x,z) & = \sum\limits_{i=-4}^0 c_3^{(i)}(x,z) \epsilon^i + O(\epsilon) \,.
\end{eqnarray}
After substituting them into Eq.~(\ref{eq:DE}) with $k=x,z$ respectively, the differential equations for $c_0^{(i)}(x,z)$ are simple first-order equations and the other three coefficients form a coupled system as expected. For simplicity, here we only show the differential equation for $c_{0,1,3}^{(-4)}$ which come with the highest pole of $\epsilon$ at the level of two loops. We get the following equations for $c_{0}^{(-4)}$, 
\begin{eqnarray}
    \frac{\partial c_0^{(-4)}(x,z)}{\partial z}  &=&  \frac{-1}{z} c_0^{(-4)}(x,z) \,, \nonumber\\
    \frac{\partial c_0^{(-4)}(x,z)}{\partial x}  &=&  \frac{2\left(1-x\right)}{x\left(1+x\right)} c_0^{(-4)}(x,z) \,.
\end{eqnarray}
Equations involving $c_{1,3}^{(-4)}$ are more complicated and we have  

{
 \footnotesize
\begin{eqnarray}
    \frac{\partial c_1^{(-4)}(x,z)}{\partial z} &=& \frac{\left(x^3 z-x^4+x z-1\right) c_1^{(-4)}(x,z)}{\left(x^2+1\right) \left(x-z\right) \left(xz-1\right)} -\frac{x \left(x^2-2 x z+1\right) c_3^{(-4)}(x,z)}{\left(x^2+1\right) \left(x-z\right) \left(xz-1\right)} \, ,\nonumber\\
    \frac{\partial c_1^{(-4)}(x,z)}{\partial x} &=&\frac{(x-1) \left(2 x \left(x^2+1\right) z^2-2 \left(x^2+1\right)^2 z+x (x-1)^2\right) c_1^{(-4)}(x,z)}{x (x+1) \left(x^2+1\right) (x-z) (x z-1)} \nonumber\\
    && +\frac{\left(1-x^2\right) z c_3^{(-4)}(x,z)}{\left(x^2+1\right) (x-z) (x z-1)} \, ,\\
    \frac{\partial c_3^{(-4)}(x,z)}{\partial z} &=& \frac{\left(x z-x^2-1\right) (x (x z-2)+z) c_3^{(-4)}(x,z)}{\left(x^2+1\right) z (x-z) (x z-1)} + \frac{\left(x^4-\left(x^2+1\right) x z+1\right) c_1^{(-4)}(x,z)}{\left(x^2+1\right) z (x-z) (x z-1)} \,, \nonumber\\
    \frac{\partial c_3^{(-4)}(x,z)}{\partial x} &=&  \frac{(1-x) \left(-2 \left(x^3+x\right)+\left(x^2+1\right) (x-1)^2 z-x (x-1)^2 z^2\right) c_3^{(-4)}(x,z)}{x (x+1) \left(x^2+1\right) (x-z) (x z-1)} \nonumber\\ 
    &&  +\frac{\left(x^4-x^3 z+x z-1\right)c_1^{(-4)}(x,z)}{x \left(x^2+1\right) (x-z) (x z-1)}  \,.
\end{eqnarray}
}
Solving all the differential equations gives us
\begin{equation}
    c^{(i)}_j(x,z) = f^{(i)}_j(x,z) + c^{(i)}_j, 
\end{equation}
where $f^{(i)}_j(x,z)$ contains all the information of MPLs and $c^{(i)}_j$ are the corresponding boundary constants.

Analytical form of $c^{(i)}_{1,2,3}$ is easy to fit since they have no relations with the bottom quark mass. Then we find that $c_0^{(-4,-3,-2,-1)}$ could be obtained without using PSLQ algorithm, 
since no MPLs need to be evaluated for them and the corresponding $\epsilon$ expanded terms of the 57th master integral vanish. On the other hand the precision problem cannot be avoided when computing $c_0^{(0)}$. Finally, we get      
\begin{eqnarray}
    c_0^{(-4)}&=&1 \,, \quad c_0^{(-3)}=0 \,, \quad c_0^{(-2)}=\frac{\pi^2}{2} \,, \quad c_0^{(-1)}=7 \zeta (3)+\frac{i \pi ^3}{2}\,, \nonumber\\
    c_0^{(0)}&=&~29.2227273102007311709320998066115333749182757 \nonumber \\
             &&-i18.8818656808153946360173789662627141075061518 
    \nonumber \\
             &=& \frac{3 \pi ^4}{10}-5 i \pi  \zeta (3).
\end{eqnarray}

At the end of this section, we would like to make some comments on the crosschecks we have made after obtaining the analytical expressions of the MIs. First, we assign the variables $x,\,z$ and $\kappa$ different groups of values and then compare the analytical results with the numerical ones computed by {\tt AMFlow}. These two results should be identical within a certain accuracy. The second method is choosing another basis of MIs and using the IBP relations to get their analytical form. Then we could check the differential equations Eq.~(\ref{eq:DE}) for this new basis. With these crosschecks we believe that our results are reliable and they are provided in the submitted ancillary file.

\section{Conclusion}
\label{Sec:con}
In this paper we computed the two-loop planar master integrals for $gg\rightarrow HH$ with an expansion in the bottom quark mass. 
The master integrals have been expanded to a sufficient order in $m_b$ to ensure that each scalar integral in the amplitude is expanded to at least $m_b^2$ order, which should be enough for phenomenological applications. Meanwhile, we maintained the full dependence on other scales at each order in $m_b$ during the calculation.
Higher order terms in $m_b$ could be easily obtained from current results through solving differential equations with respect to the quark mass.  
We want to stress that in order to get rid of the possible errors caused by the numerical method used for the determination of boundary constants, we demanded that stable results should be obtained by the PSQL algorithm. 
To ensure this stability, we sometimes even incorporated the contributions of higher order terms in $m_b$ for specific master integrals.
Furthermore, we calculated the MIs at other physical points of phase space and chose a different basis in each integral family to check our analytical expressions. Based on the experience from this work, we believe that the non-planar Feynman integrals could also be solved in the same way, and they will be considered in our next work.

\acknowledgments
We are grateful to Yanghua Guo and Shiqi Wang for collaboration at early stage of this project. We would like to thank Yang Zhang for helpful discussions on the property of MPLs. 
This work is supported in part by the National Natural Science Foundation of China (NNSFC) under grant Nos. 12375082, 12135013 and 12335005, by Institute of High Energy Physics (IHEP, CAS) under Grants No. Y9515570U1, and by the Research Fund (Grant No. 5101029470335) from Henan Normal University.
\newpage

\appendix

\section{Integral families and master integrals}
\label{appendixA}

For $g(p_1)g(p_2) \rightarrow H(p_3) H(p_4)$ with all momenta incoming, we define the one-loop integral families as
\begin{align}
    \text{B1} = \left\{
k_1^2-m_b^2\,,
(k_1+p_1)^2-m_b^2\,,
(k_1+p_1+p_2)^2-m_b^2\,,
(k_1+p_1+p_2+p_3)^2-m_b^2\,
\right\}\,,\nonumber
\end{align}
where $k_1$ is the loop momentum and $\text{B1x23} = \text{B1}(p_2 \leftrightarrow p_3)$. The master integrals are
\begin{align}
    \begin{array}{llll}
    \text{B1}(1,0,0,0), & \text{B1}(1,0,1,0), & \text{B1}(1,0,0,1), & \text{B1}(0,1,0,1), \\
    \text{B1}(1,1,1,0), & \text{B1}(1,1,0,1), & \text{B1}(1,0,1,1), & \text{B1}(1,1,1,1), \\
    \text{B1x23}(1,0,1,0), & \text{B1x23}(1,1,1,0), & \text{B1x23}(1,1,1,1). \nonumber
    \end{array}
\end{align}
We have compared our results with those computed by \emph{Package}-{\tt\bfseries X}~\cite{Patel:2016fam} and found that all results are identical except which belong to four-point MIs. This discrepancy comes from a bug in {\tt D0Expand} functions when $\kappa$ is small.\footnote{Here we used the function {\tt D0Expand} in \emph{Package}-{\tt\bfseries X} to get analytical expressions, while it was found that {\tt D0Expand} could produce numerically different results comparing with the original {\tt D0} function.} Thus, instead of directly comparing four-point MIs, $x$ or $z$ was fixed at different values and the four-point functions were reduced to three-point ones. Then agreements were found between them for several different values of $x$ or $z$.

At two-loop level, we have 10 planar integral families which are defined as
\begin{align}
\text{PL1} & = \left\{
k_1^2\,,
k_2^2-m_b^2\,,
(k_1-k_2)^2-m_b^2\,,
(k_1-p_1)^2\,,
(k_1-p_1-p_2)^2\,,
(k_2-p_1-p_2)^2-m_b^2\,,
\right.\nonumber\\&\left.
(k_2-p_1-p_2-p_3)^2-m_b^2\,,
(k_2-p_1)^2\,,
(k_1-p_1-p_2-p_3)^2\,
\right\}\,,\nonumber\\
\text{PL2} & = \left\{
k_1^2-m_b^2\,,
k_2^2-m_b^2\,,
(k_1-k_2)^2\,,
(k_2-p_1-p_2-p_3)^2-m_b^2\,,
\right.\nonumber\\&\left.
(k_1-p_1-p_2-p_3)^2-m_b^2\,,
(k_1-p_1-p_2)^2-m_b^2\,,
(k_1-p_1)^2-m_b^2\,,
(k_2-p_1)^2\,,
\right.\nonumber\\&\left.
(k_2-p_1-p_2)^2\,
\right\}\,,\nonumber\\
\text{PL3} & = \left\{
k_1^2-m_b^2\,,
k_2^2-m_b^2\,,
(k_1-k_2)^2\,,
(k_2-p_1-p_2-p_3)^2-m_b^2\,,
(k_2-p_1-p_2)^2-m_b^2\,,
\right.\nonumber\\&\left.
(k_2-p_1)^2-m_b^2\,,
(k_1-p_1)^2-m_b^2\,,
(k_1-p_1-p_2)^2\,,
(k_1-p_1-p_2-p_3)^2\,
\right\}\,,\nonumber\\
\text{PL4} & = \left\{
k_1^2\,,
k_2^2-m_b^2\,,
(k_1-k_2)^2-m_b^2\,,
(k_2-p_1-p_2-p_3)^2-m_b^2\,,
(k_2-p_1-p_2)^2-m_b^2\,,
\right.\nonumber\\&\left.
(k_2-p_1)^2-m_b^2\,,
(k_1-p_1)^2\,,
(k_1-p_1-p_2)^2\,,
(k_1-p_1-p_2-p_3)^2\,
\right\}\,,\nonumber\\
\text{PL5} & = \left\{
k_1^2-m_b^2\,,
k_2^2-m_b^2\,,
(k_1-k_2)^2\,,
(k_1-p_1)^2-m_b^2\,,
(k_1-p_1-p_2)^2-m_b^2\,,
\right.\nonumber\\&\left.
(k_2-p_1-p_2)^2-m_b^2\,,
(k_2-p_1-p_2-p_3)^2-m_b^2\,,
(k_2-p_1)^2\,,
(k_1-p_1-p_2-p_3)^2\,
\right\}\,,\nonumber\\
\text{PL6} & = \left\{
k_1^2-m_b^2\,,
(k_1-k_2)^2\,,
(k_1-p_1)^2-m_b^2\,,
(k_2-p_1)^2-m_b^2\,,
(k_2-p_1-p_2)^2-m_b^2\,,
\right.\nonumber\\&\left.
(k_1-p_1-p_2-p_3)^2-m_b^2\,,
(k_2-p_1-p_2-p_3)^2-m_b^2\,,
k_2^2\,,
(k_1-p_1-p_2)^2\,
\right\}\,.
\nonumber
\end{align}
Here $k_1,k_2$ are the loop momenta. The other four families are obtained by the following permutations,
\begin{align}
\begin{array}{cc}
   \text{PL2x23}  = \text{PL2}(p_2 \leftrightarrow p_3)\,, \quad & \text{PL3x23}  = \text{PL3}(p_2 \leftrightarrow p_3)\,,\\
    \text{PL4x23}  = \text{PL4}(p_2 \leftrightarrow p_3)\,, \quad & 
    \text{PL5x23}  = \text{PL5}(p_2 \leftrightarrow p_3)\,.\nonumber
\end{array}
\end{align}
After reduction, we get 177 master integrals which are
{
\tiny
\begin{align}
\begin{array}{llll} 
   \text{PL1}(0,1,1,0,0,0,0,0,0),&\text{PL1}(1,1,0,0,1,0,0,0,0),&\text{PL1}(0,1,1,0,0,1,0,0,0),&\text{PL1}(0,1,1,0,0,0,1,0,0),\\
   \text{PL1}(1,0,1,0,0,0,1,0,0),&\text{PL1}(1,0,1,0,0,0,2,0,0),&\text{PL1}(0,1,1,0,1,0,0,0,0),&\text{PL1}(0,1,1,0,2,0,0,0,0),\\
   \text{PL1}(0,0,1,1,0,0,1,0,0),&\text{PL1}(0,0,1,1,0,0,2,0,0),&\text{PL1}(1,1,0,0,1,1,0,0,0),&\text{PL1}(1,1,0,0,1,0,1,0,0),\\
   \text{PL1}(0,1,1,0,0,1,1,0,0),&\text{PL1}(1,0,1,0,1,0,1,0,0),&\text{PL1}(1,0,1,0,2,0,1,0,0),&\text{PL1}(1,0,1,0,1,0,2,0,0),\\
   \text{PL1}(0,1,1,1,0,1,0,0,0),&\text{PL1}(0,1,1,2,0,1,0,0,0),&\text{PL1}(0,1,1,1,0,2,0,0,0),&\text{PL1}(0,1,1,1,0,0,1,0,0),\\
   \text{PL1}(0,1,1,2,0,0,1,0,0),&\text{PL1}(0,1,1,1,0,0,2,0,0),&\text{PL1}(0,1,1,0,1,0,1,0,0),&\text{PL1}(0,1,1,0,2,0,1,0,0),\\
   \text{PL1}(0,1,1,0,1,0,2,0,0),&\text{PL1}(1,1,1,0,1,1,0,0,0),&\text{PL1}(1,1,0,0,1,1,1,0,0),&\text{PL1}(1,1,1,0,1,0,1,0,0),\\
   \text{PL1}(1,1,2,0,1,0,1,0,0),&\text{PL1}(1,1,1,0,1,0,2,0,0),&\text{PL1}(1,0,1,1,1,0,1,0,0),&\text{PL1}(1,0,1,1,1,0,2,0,0),\\
   \text{PL1}(0,1,1,1,1,0,1,0,0),&\text{PL1}(0,1,1,1,2,0,1,0,0),&\text{PL1}(0,1,1,1,1,0,2,0,0),&\text{PL1}(0,1,1,1,0,1,1,0,0),\\
   \text{PL1}(0,1,1,1,0,2,1,0,0),&\text{PL1}(0,1,1,1,0,1,2,0,0),&\text{PL1}(1,1,1,0,1,1,1,0,0),&\text{PL1}(1,1,1,1,1,0,1,0,0),\\
   \text{PL1}(1,1,1,1,1,0,2,0,0),&\text{PL1}(1,1,1,1,1,1,1,0,0),&\text{PL1}(1,1,1,2,1,1,1,0,0),&\text{PL1}(1,1,1,1,1,2,1,0,0),\\
   \text{PL1}(1,1,1,1,1,1,2,0,0),&\text{PL2}(0,1,0,0,1,0,1,0,0),&\text{PL2}(1,1,0,1,1,0,0,0,0),&\text{PL2}(1,1,0,1,0,1,0,0,0),\\
   \text{PL2}(1,0,1,1,0,1,0,0,0),&\text{PL2}(1,0,1,1,0,2,0,0,0),&\text{PL2}(1,0,1,2,0,1,0,0,0),&\text{PL2}(1,0,1,1,0,0,1,0,0),\\
   \text{PL2}(1,1,0,0,1,0,1,0,0),&\text{PL2}(0,1,1,0,1,0,1,0,0),&\text{PL2}(0,1,1,0,1,0,2,0,0),&\text{PL2}(0,1,1,0,2,0,1,0,0),\\
   \text{PL2}(0,1,0,1,1,0,1,0,0),&\text{PL2}(1,1,0,0,0,1,1,0,0),&\text{PL2}(0,1,1,0,0,1,1,0,0),&\text{PL2}(1,1,1,1,0,1,0,0,0),\\
   \text{PL2}(1,1,1,1,0,2,0,0,0),&\text{PL2}(1,1,1,2,0,1,0,0,0),&\text{PL2}(1,1,2,1,0,1,0,0,0),&\text{PL2}(1,1,0,1,1,1,0,0,0),\\
   \text{PL2}(0,1,1,1,1,1,0,0,0),&\text{PL2}(0,1,1,1,1,2,0,0,0),&\text{PL2}(0,1,2,1,1,1,0,0,0),&\text{PL2}(1,1,1,1,0,0,1,0,0),\\
   \text{PL2}(1,1,1,1,0,0,2,0,0),&\text{PL2}(1,1,0,1,1,0,1,0,0),&\text{PL2}(0,1,1,1,1,0,1,0,0),&\text{PL2}(0,1,1,1,1,0,2,0,0),\\
   \text{PL2}(0,1,1,2,1,0,1,0,0),&\text{PL2}(0,1,2,1,1,0,1,0,0),&\text{PL2}(1,1,0,1,0,1,1,0,0),&\text{PL2}(1,0,1,1,0,1,1,0,0),\\
   \text{PL2}(1,0,1,1,0,1,2,0,0),&\text{PL2}(1,0,1,1,0,2,1,0,0),&\text{PL2}(0,1,1,1,0,1,1,0,0),&\text{PL2}(0,1,1,1,0,1,2,0,0),\\
   \text{PL2}(0,1,1,1,0,2,1,0,0),&\text{PL2}(0,1,1,2,0,1,1,0,0),&\text{PL2}(0,1,2,1,0,1,1,0,0),&\text{PL2}(1,1,0,0,1,1,1,0,0),\\
   \text{PL2}(0,1,1,0,1,1,1,0,0),&\text{PL2}(0,1,1,0,1,1,2,0,0),&\text{PL2}(0,1,1,0,1,2,1,0,0),&\text{PL2}(1,1,1,1,0,1,1,0,0),\\
   \text{PL2}(1,1,1,1,0,1,2,0,0),&\text{PL2}(1,1,1,1,0,2,1,0,0),&\text{PL2}(1,1,1,2,0,1,1,0,0),&\text{PL2}(1,1,0,1,1,1,1,0,0),\\
   \text{PL2}(0,1,1,1,1,1,1,0,0),&\text{PL2}(0,1,1,1,1,1,2,0,0),&\text{PL2}(0,1,1,1,1,2,1,0,0),&\text{PL2}(0,1,1,2,1,1,1,0,0),\\
   \text{PL3}(1,0,1,0,1,1,1,0,0),&\text{PL3}(1,1,1,0,1,0,1,0,0),&\text{PL3}(1,1,1,0,1,0,2,0,0),&\text{PL3}(1,0,1,1,0,1,1,0,0),\\
   \text{PL3}(1,0,1,1,0,1,2,0,0),&\text{PL3}(1,0,1,1,1,1,1,0,0),&\text{PL3}(2,0,1,1,1,1,1,0,0),&\text{PL3}(1,1,1,1,1,0,1,0,0),\\
   \text{PL3}(1,1,1,1,1,0,2,0,0),&\text{PL5}(1,1,0,0,1,1,0,0,0),&\text{PL5}(1,1,0,1,1,1,0,0,0),&\text{PL5}(1,1,0,0,1,1,1,0,0),\\
   \text{PL5}(1,1,0,1,1,1,1,0,0),&\text{PL5}(1,1,1,1,1,1,1,0,0),&\text{PL5}(1,1,1,1,1,1,1,-1,0),&\text{PL5}(1,1,1,1,2,1,1,0,0),\\
   \text{PL5}(1,1,1,1,1,1,2,0,0),&\text{PL6}(0,0,1,1,0,1,1,0,0),&\text{PL6}(1,0,1,1,0,1,1,0,0),&\text{PL6}(1,0,1,1,1,1,1,0,0),\\
   \text{PL6}(1,1,1,1,1,1,1,0,0),&\text{PL6}(1,1,1,1,1,1,1,-1,0),&\text{PL6}(1,1,1,1,1,1,2,0,0),&\text{PL2x23}(1,1,0,0,0,1,0,0,0),\\
   \text{PL2x23}(0,1,1,0,0,1,0,0,0),&\text{PL2x23}(0,1,1,0,0,2,0,0,0),&\text{PL2x23}(1,1,0,1,0,1,0,0,0),&\text{PL2x23}(1,1,0,0,1,1,0,0,0),\\
   \text{PL2x23}(0,1,1,0,1,1,0,0,0),&\text{PL2x23}(1,0,1,1,0,1,0,0,0),&\text{PL2x23}(1,0,1,2,0,1,0,0,0),&\text{PL2x23}(1,0,1,1,0,2,0,0,0),\\
   \text{PL2x23}(0,1,1,1,0,1,0,0,0),&\text{PL2x23}(0,1,1,2,0,1,0,0,0),&\text{PL2x23}(0,1,1,1,0,2,0,0,0),&\text{PL2x23}(1,1,0,1,1,1,0,0,0),\\
   \text{PL2x23}(1,1,0,0,1,1,1,0,0),&\text{PL2x23}(1,1,1,1,0,1,0,0,0),&\text{PL2x23}(1,1,2,1,0,1,0,0,0),&\text{PL2x23}(1,1,1,2,0,1,0,0,0),\\
   \text{PL2x23}(1,1,1,1,0,2,0,0,0),&\text{PL2x23}(1,0,1,1,0,1,1,0,0),&\text{PL2x23}(1,0,1,1,0,2,1,0,0),&\text{PL2x23}(1,0,1,1,0,1,2,0,0),\\
   \text{PL2x23}(0,1,1,1,1,1,0,0,0),&\text{PL2x23}(0,1,1,1,1,2,0,0,0),&\text{PL2x23}(0,1,1,0,1,1,1,0,0),&\text{PL2x23}(0,1,1,0,1,2,1,0,0),\\
   \text{PL2x23}(0,1,1,0,1,1,2,0,0),&\text{PL2x23}(0,1,1,1,0,1,1,0,0),&\text{PL2x23}(0,1,2,1,0,1,1,0,0),&\text{PL2x23}(0,1,1,1,0,2,1,0,0),\\
   \text{PL2x23}(0,1,1,1,0,1,2,0,0),&\text{PL2x23}(0,1,1,1,0,1,3,0,0),&\text{PL2x23}(1,1,0,1,1,1,1,0,0),&\text{PL2x23}(1,1,1,1,0,1,1,0,0),\\
   \text{PL2x23}(1,1,1,2,0,1,1,0,0),&\text{PL2x23}(1,1,1,1,0,2,1,0,0),&\text{PL2x23}(1,1,1,1,0,1,2,0,0),&\text{PL2x23}(0,1,1,1,1,1,1,0,0),\\
   \text{PL2x23}(0,1,1,2,1,1,1,0,0),&\text{PL2x23}(0,1,1,1,1,2,1,0,0),&\text{PL2x23}(0,1,1,1,1,1,2,0,0),&\text{PL3x23}(1,1,1,0,1,0,1,0,0),\\
   \text{PL3x23}(1,1,1,0,1,0,2,0,0),&\text{PL3x23}(1,0,1,1,1,0,1,0,0),&\text{PL3x23}(1,0,2,1,1,0,1,0,0),&\text{PL3x23}(1,0,1,1,1,0,2,0,0),\\
   \text{PL3x23}(1,0,1,1,1,1,1,0,0),&\text{PL3x23}(2,0,1,1,1,1,1,0,0),&\text{PL3x23}(1,1,1,1,1,0,1,0,0),&\text{PL3x23}(1,1,1,1,1,0,2,0,0),\\
   \text{PL4x23}(1,0,1,1,1,0,1,0,0),&\text{PL4x23}(1,0,1,1,2,0,1,0,0),&\text{PL4x23}(1,0,1,1,1,0,2,0,0),&\text{PL5x23}(1,1,0,0,1,1,0,0,0),\\
   \text{PL5x23}(1,1,0,1,1,1,0,0,0),&\text{PL5x23}(1,1,0,1,1,1,1,0,0),&\text{PL5x23}(1,1,1,1,1,1,1,0,0),&\text{PL5x23}(1,1,2,1,1,1,1,0,0),\\
   \text{PL5x23}(1,1,1,1,2,1,1,0,0).
   \nonumber
\end{array}
\end{align}
}
Each family is defined with nine propagators, the first seven ones of which carry positive powers and the others own non-positive powers. We prioritize choosing the integrals without negative powers and at the same time with relatively small positive power for each propagator as the master integrals of this family.
For convenience we show some graphical representations of these two-loop master integrals in Fig.~\ref{Fig:MIs}.

\newpage

\begin{figure}[H]
    \centering 
	\begin{minipage}{0.16\textwidth} 
		\centering 
		\includegraphics[width=\textwidth]{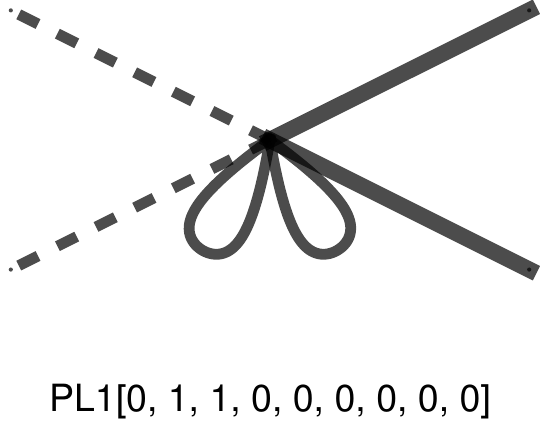} 
	\end{minipage}
    \begin{minipage}{0.16\textwidth} 
		\centering 
		\includegraphics[width=\textwidth]{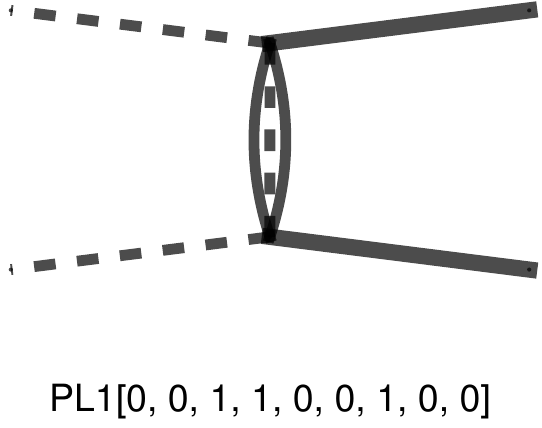} 
	\end{minipage}
    \begin{minipage}{0.16\textwidth} 
		\centering 
		\includegraphics[width=\textwidth]{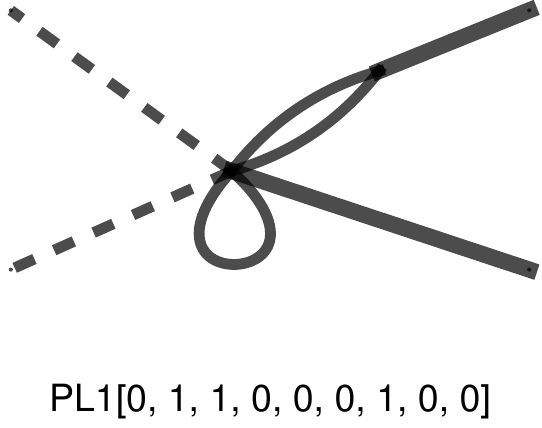} 
	\end{minipage}
	\begin{minipage}{0.16\textwidth} 
		\centering 
		\includegraphics[width=\textwidth]{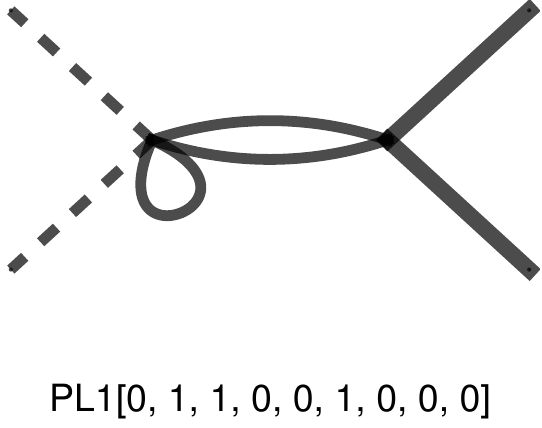} 
	\end{minipage}
	\begin{minipage}{0.16\textwidth} 
		\centering 
		\includegraphics[width=\textwidth]{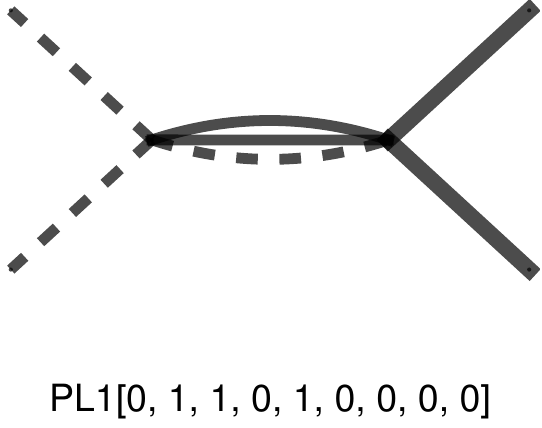} 
	\end{minipage}
	\begin{minipage}{0.16\textwidth} 
		\centering 
		\includegraphics[width=\textwidth]{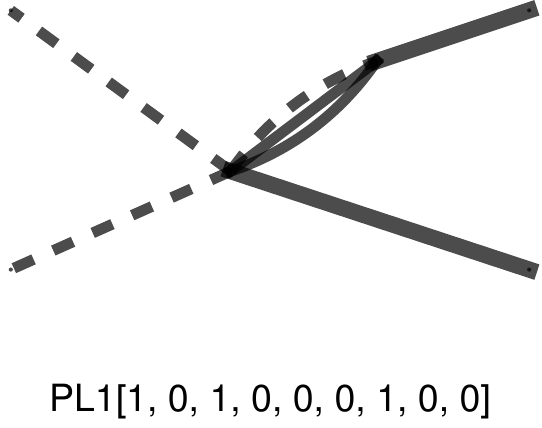} 
	\end{minipage}
    
    \begin{minipage}{0.16\textwidth} 
		\centering 
		\includegraphics[width=\textwidth]{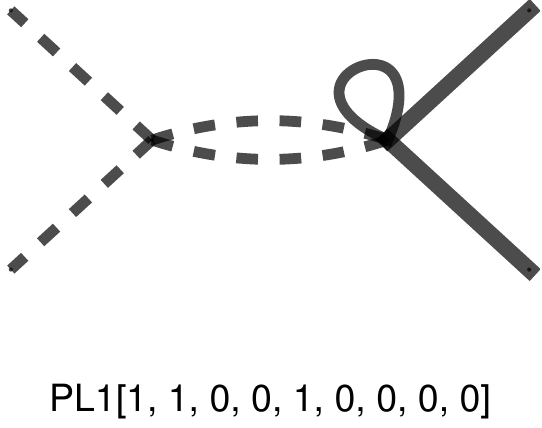} 
	\end{minipage}
    \begin{minipage}{0.16\textwidth} 
		\centering 
		\includegraphics[width=\textwidth]{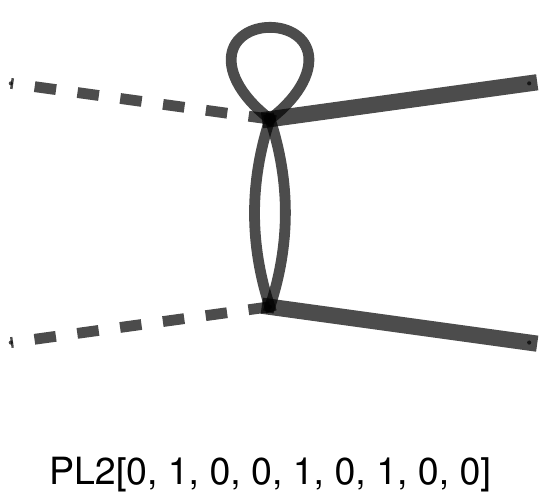} 
	\end{minipage}
    \begin{minipage}{0.16\textwidth} 
		\centering 
		\includegraphics[width=\textwidth]{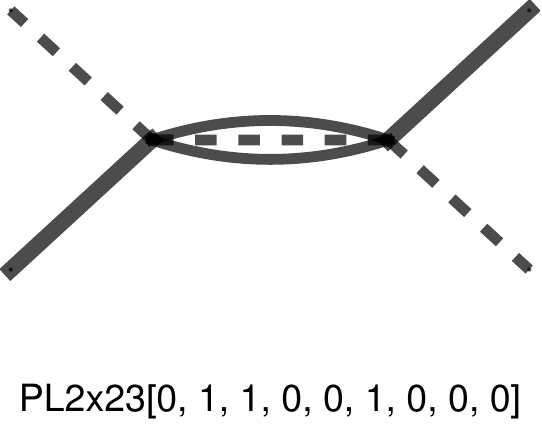} 
	\end{minipage}
	\begin{minipage}{0.16\textwidth} 
		\centering 
		\includegraphics[width=\textwidth]{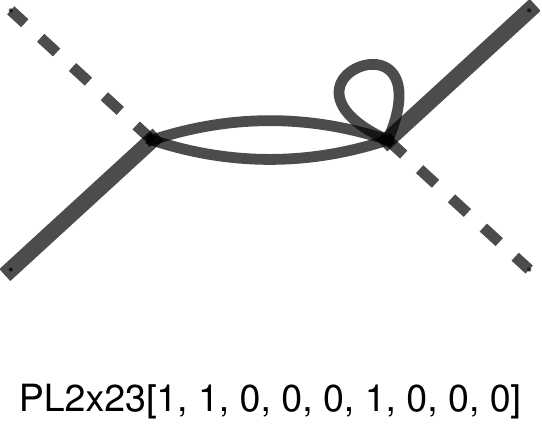} 
	\end{minipage}
	\begin{minipage}{0.16\textwidth} 
		\centering 
		\includegraphics[width=\textwidth]{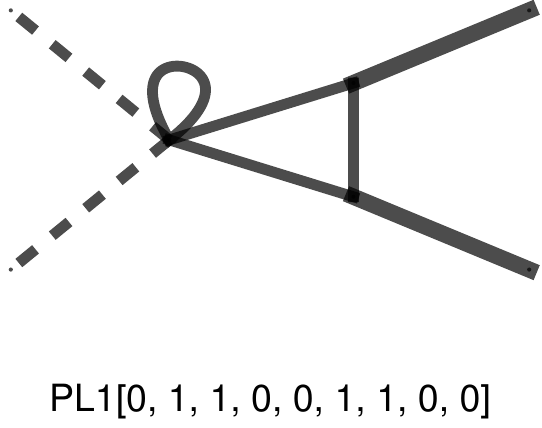} 
	\end{minipage}
	\begin{minipage}{0.16\textwidth} 
		\centering 
		\includegraphics[width=\textwidth]{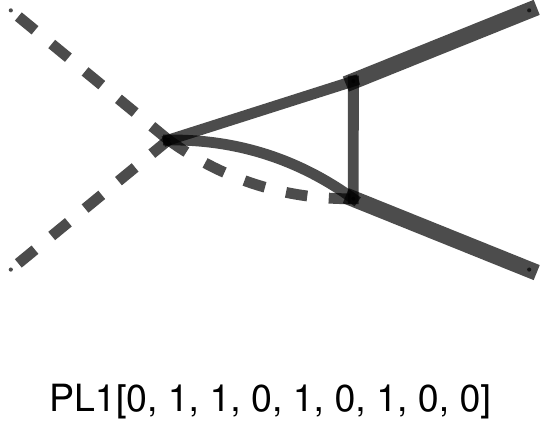} 
	\end{minipage}
    
    \begin{minipage}{0.16\textwidth} 
		\centering 
		\includegraphics[width=\textwidth]{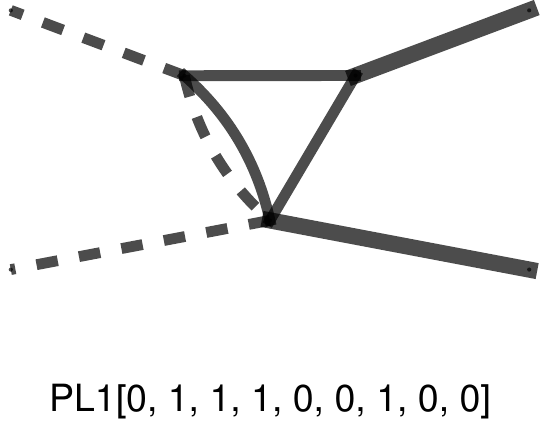} 
	\end{minipage}
    \begin{minipage}{0.16\textwidth} 
		\centering 
		\includegraphics[width=\textwidth]{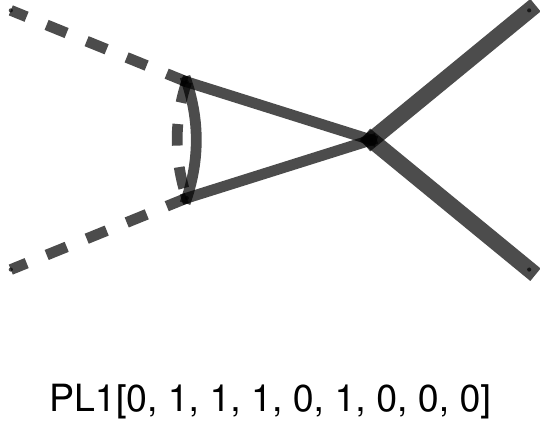} 
	\end{minipage}
    \begin{minipage}{0.16\textwidth} 
		\centering 
		\includegraphics[width=\textwidth]{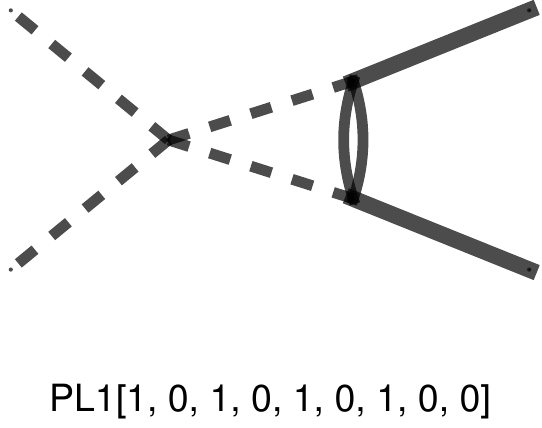} 
	\end{minipage}
	\begin{minipage}{0.16\textwidth} 
		\centering 
		\includegraphics[width=\textwidth]{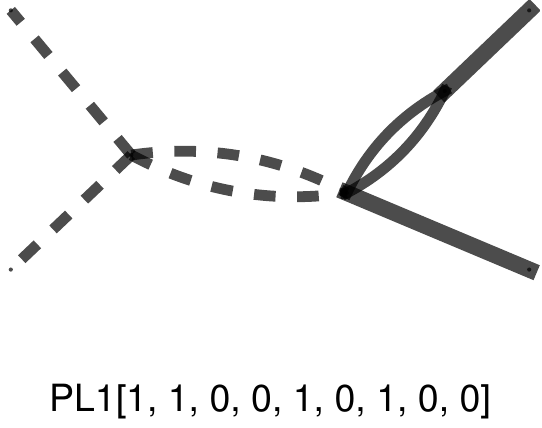} 
	\end{minipage}
    \begin{minipage}{0.16\textwidth} 
		\centering 
		\includegraphics[width=\textwidth]{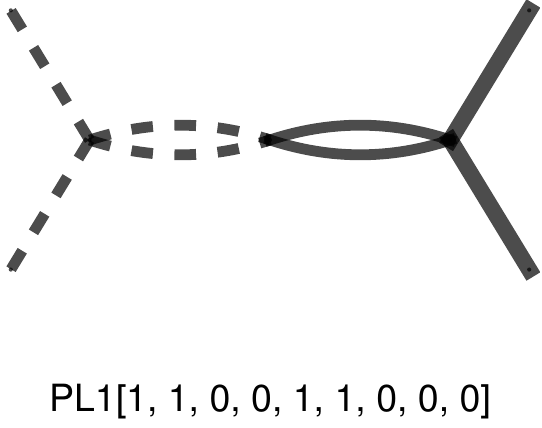} 
	\end{minipage}
	\begin{minipage}{0.16\textwidth} 
		\centering 
		\includegraphics[width=\textwidth]{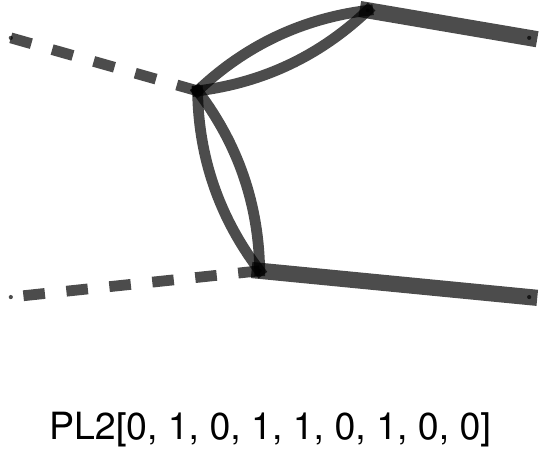} 
	\end{minipage}
    
    \begin{minipage}{0.16\textwidth} 
		\centering 
		\includegraphics[width=\textwidth]{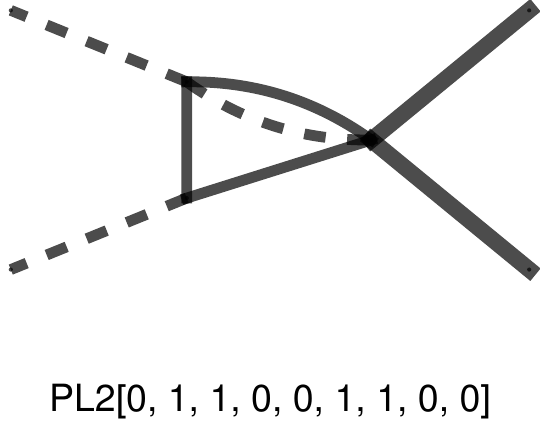} 
	\end{minipage}
    \begin{minipage}{0.16\textwidth} 
		\centering 
		\includegraphics[width=\textwidth]{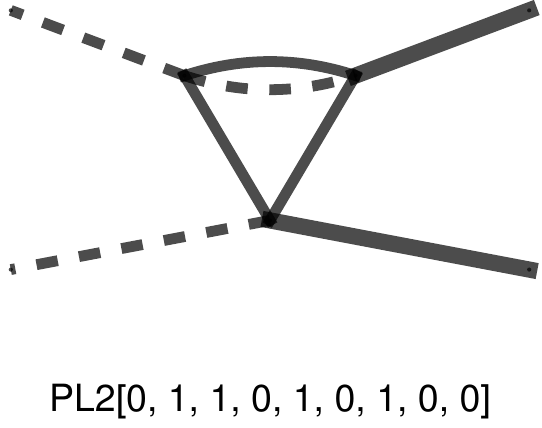} 
	\end{minipage}
    \begin{minipage}{0.16\textwidth} 
		\centering 
		\includegraphics[width=\textwidth]{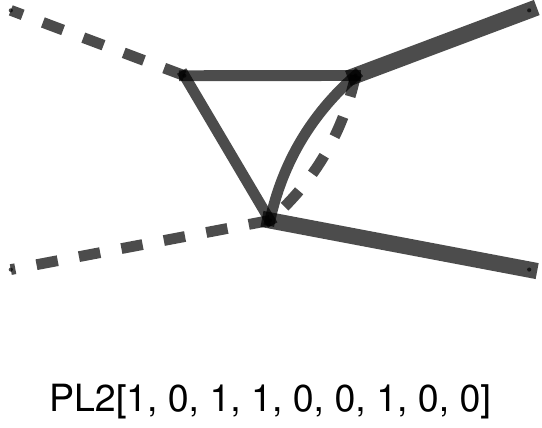} 
	\end{minipage}
	\begin{minipage}{0.16\textwidth} 
		\centering 
		\includegraphics[width=\textwidth]{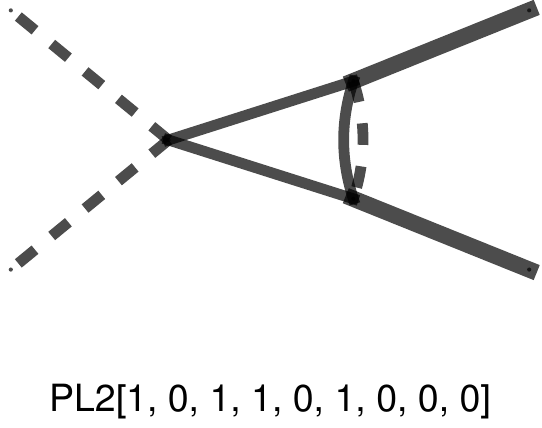} 
	\end{minipage}
	\begin{minipage}{0.16\textwidth} 
		\centering 
		\includegraphics[width=\textwidth]{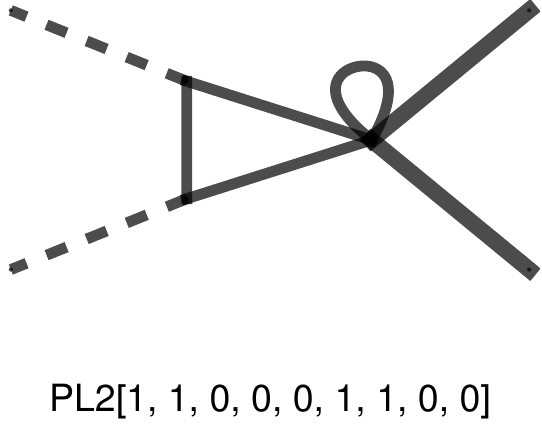} 
	\end{minipage}
	\begin{minipage}{0.16\textwidth} 
		\centering 
		\includegraphics[width=\textwidth]{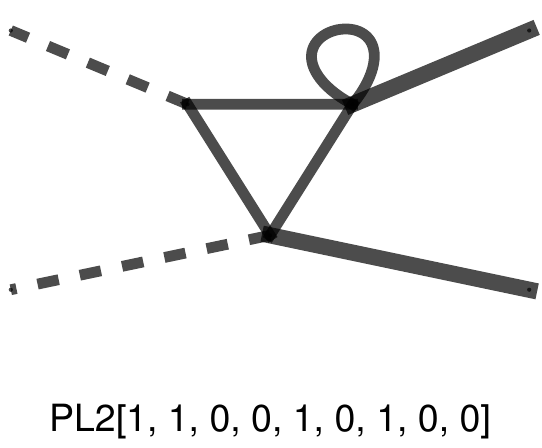} 
	\end{minipage}

    \begin{minipage}{0.16\textwidth} 
		\centering 
		\includegraphics[width=\textwidth]{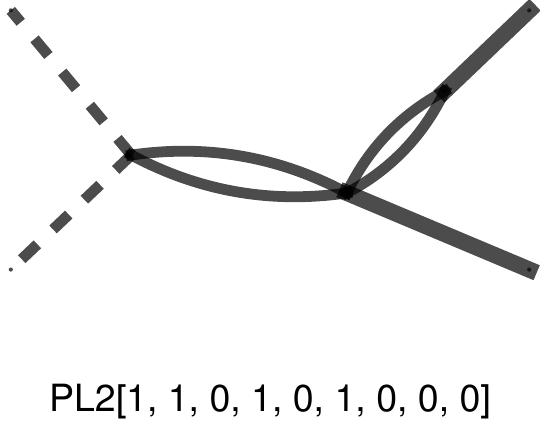} 
	\end{minipage}
    \begin{minipage}{0.16\textwidth} 
		\centering 
		\includegraphics[width=\textwidth]{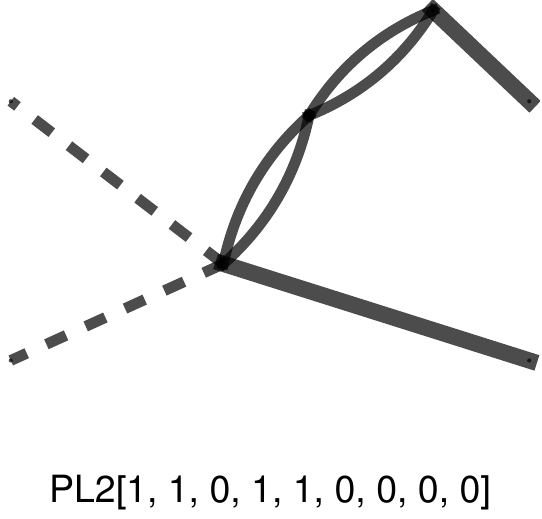} 
	\end{minipage}
    \begin{minipage}{0.16\textwidth} 
		\centering 
		\includegraphics[width=\textwidth]{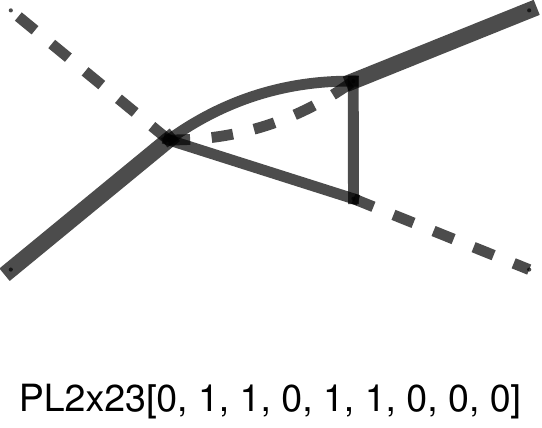} 
	\end{minipage}
	\begin{minipage}{0.16\textwidth} 
		\centering 
		\includegraphics[width=\textwidth]{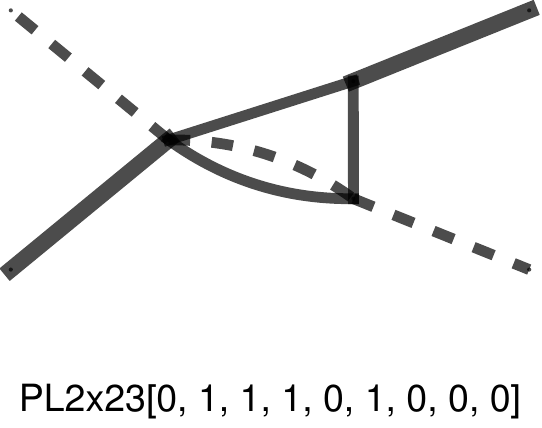} 
	\end{minipage}
	\begin{minipage}{0.16\textwidth} 
		\centering 
		\includegraphics[width=\textwidth]{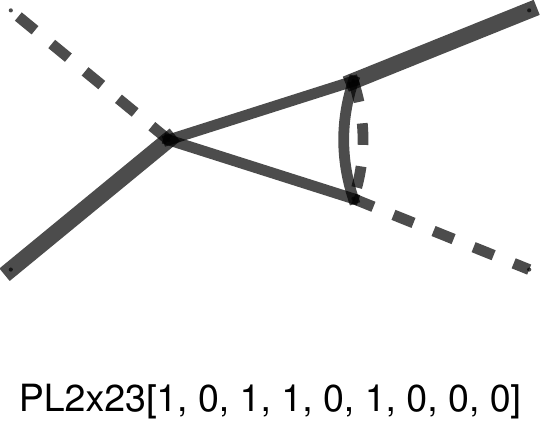} 
	\end{minipage}
	\begin{minipage}{0.16\textwidth} 
		\centering 
		\includegraphics[width=\textwidth]{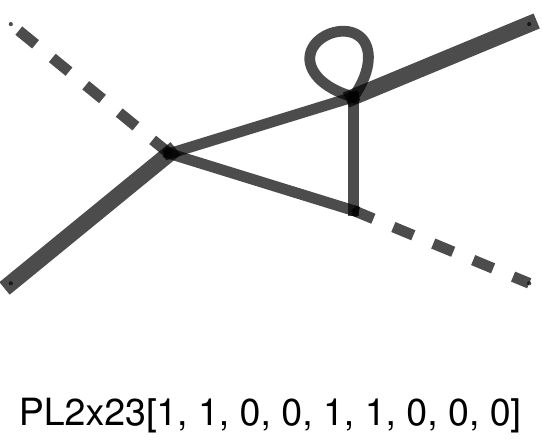} 
	\end{minipage}
    
    \begin{minipage}{0.16\textwidth} 
		\centering 
		\includegraphics[width=\textwidth]{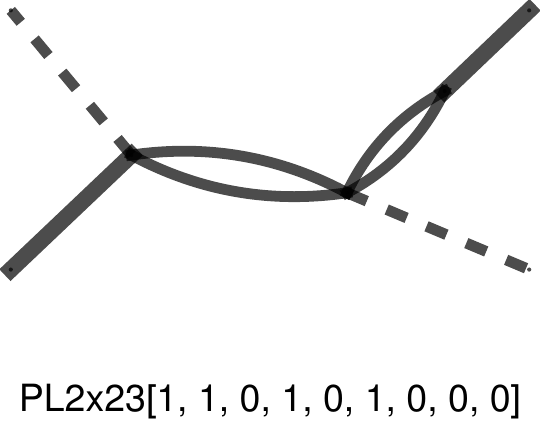} 
	\end{minipage}
    \begin{minipage}{0.16\textwidth} 
		\centering 
		\includegraphics[width=\textwidth]{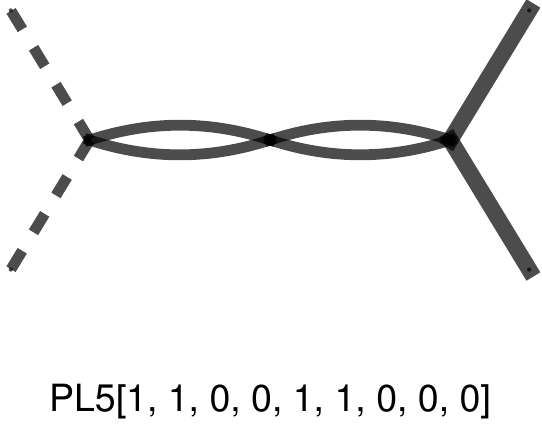} 
	\end{minipage}
    \begin{minipage}{0.16\textwidth} 
		\centering 
		\includegraphics[width=\textwidth]{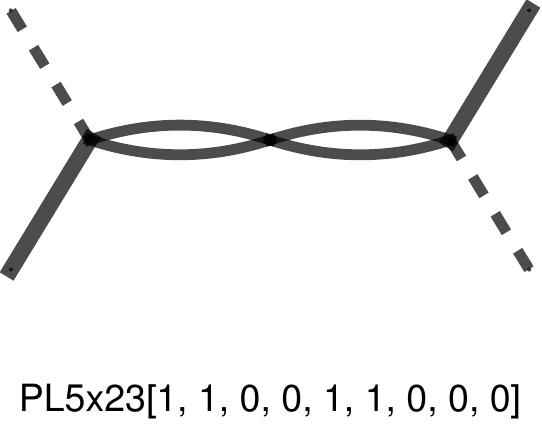} 
	\end{minipage}
	\begin{minipage}{0.16\textwidth} 
		\centering 
		\includegraphics[width=\textwidth]{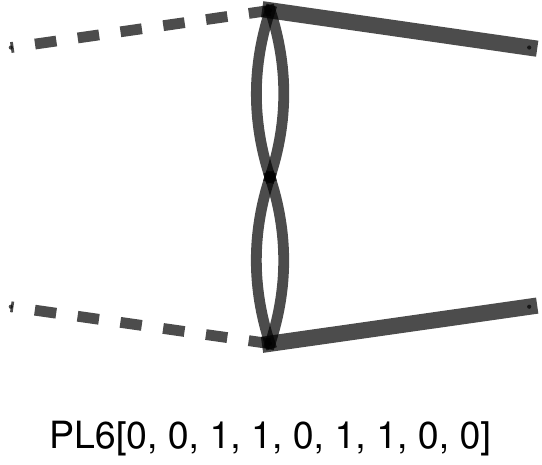} 
	\end{minipage}
	\begin{minipage}{0.16\textwidth} 
		\centering 
		\includegraphics[width=\textwidth]{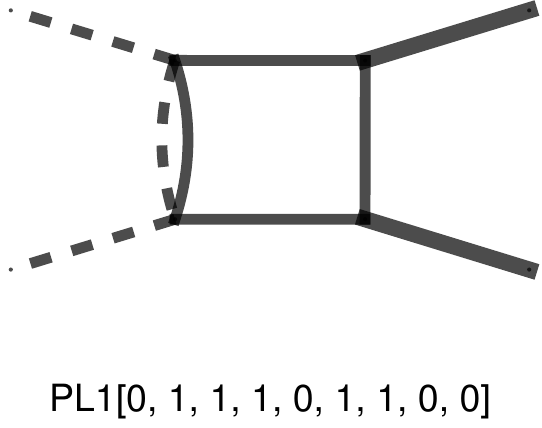} 
	\end{minipage}
	\begin{minipage}{0.16\textwidth} 
		\centering 
		\includegraphics[width=\textwidth]{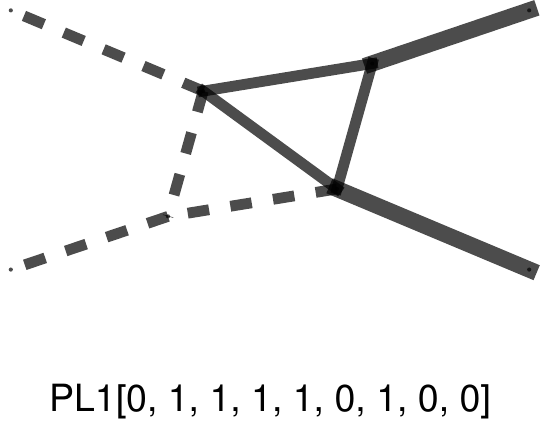} 
	\end{minipage}
    
    \begin{minipage}{0.16\textwidth} 
		\centering 
		\includegraphics[width=\textwidth]{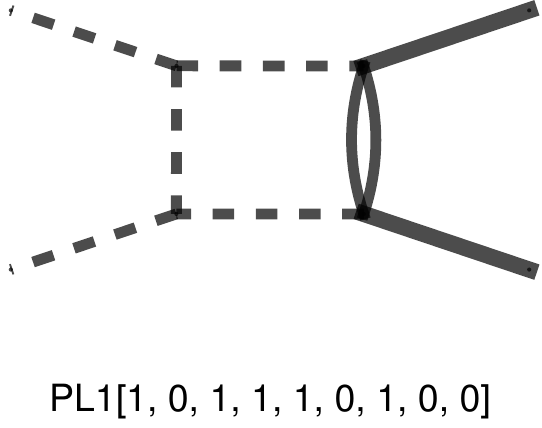} 
	\end{minipage}
    \begin{minipage}{0.16\textwidth} 
		\centering 
		\includegraphics[width=\textwidth]{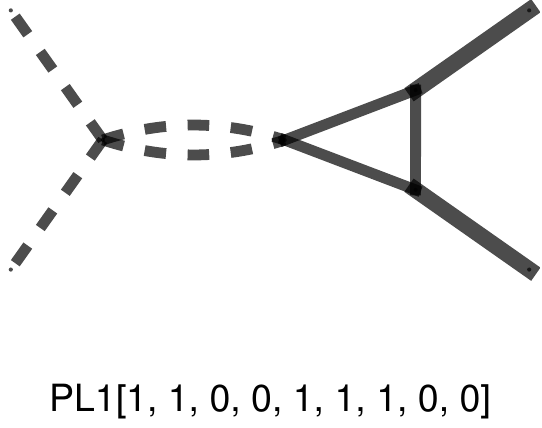} 
	\end{minipage}
    \begin{minipage}{0.16\textwidth} 
		\centering 
		\includegraphics[width=\textwidth]{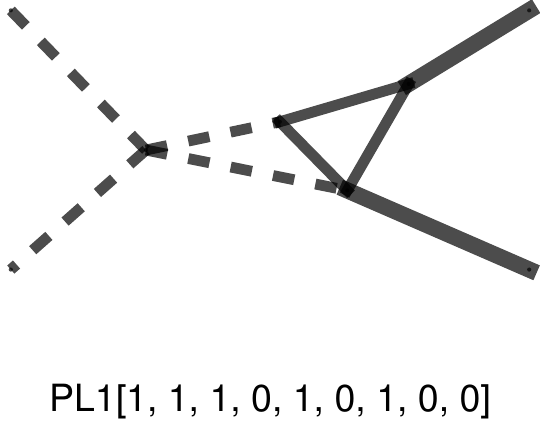} 
	\end{minipage}
	\begin{minipage}{0.16\textwidth} 
		\centering 
		\includegraphics[width=\textwidth]{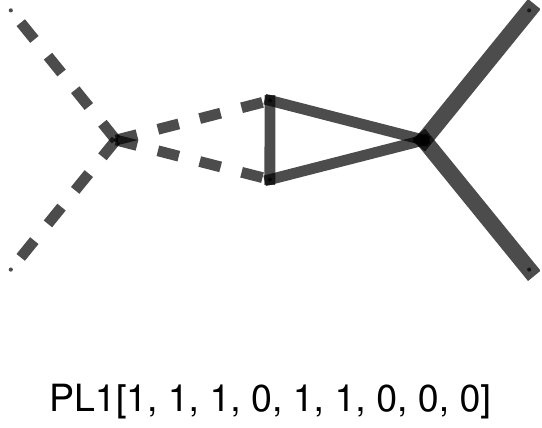} 
	\end{minipage}
    \begin{minipage}{0.16\textwidth} 
		\centering 
		\includegraphics[width=\textwidth]{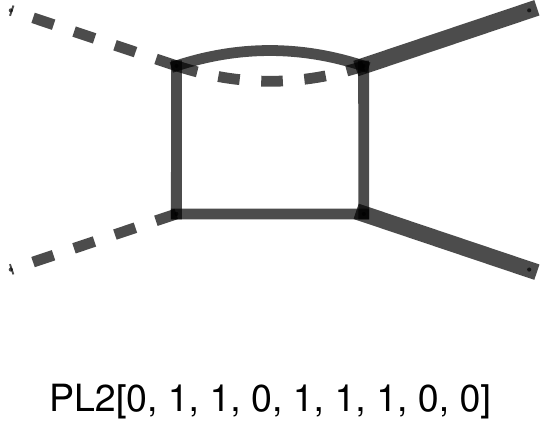} 
	\end{minipage}
	\begin{minipage}{0.16\textwidth} 
		\centering 
		\includegraphics[width=\textwidth]{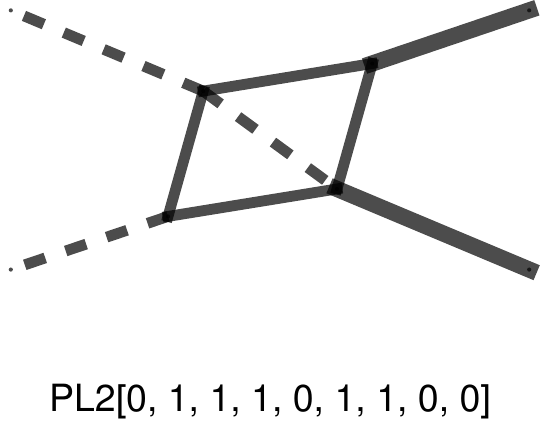} 
	\end{minipage}
    
    \begin{minipage}{0.16\textwidth} 
		\centering 
		\includegraphics[width=\textwidth]{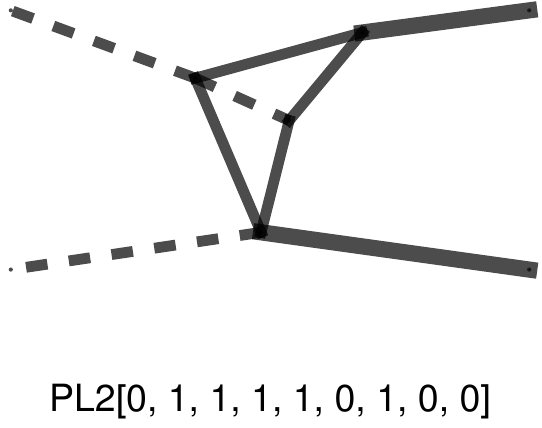} 
	\end{minipage}
    \begin{minipage}{0.16\textwidth} 
		\centering 
		\includegraphics[width=\textwidth]{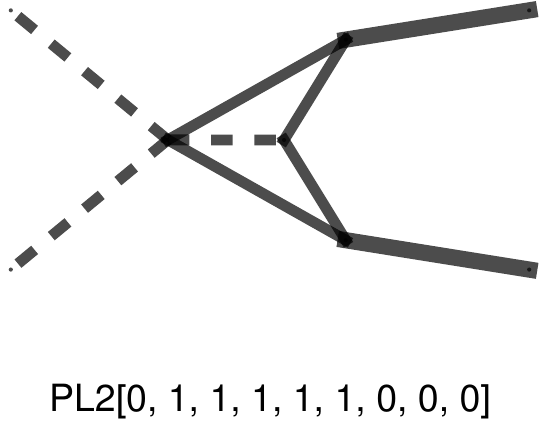} 
	\end{minipage}
    \begin{minipage}{0.16\textwidth} 
		\centering 
		\includegraphics[width=\textwidth]{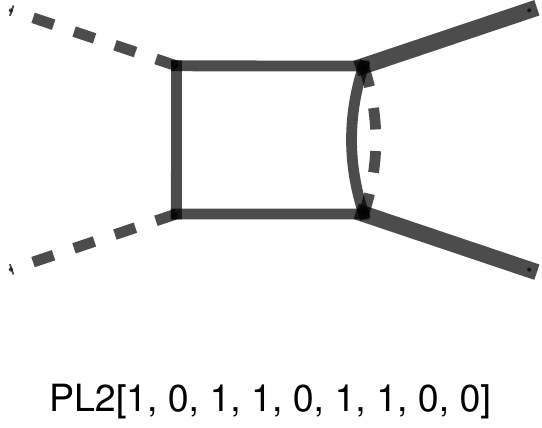} 
	\end{minipage}
	\begin{minipage}{0.16\textwidth} 
		\centering 
		\includegraphics[width=\textwidth]{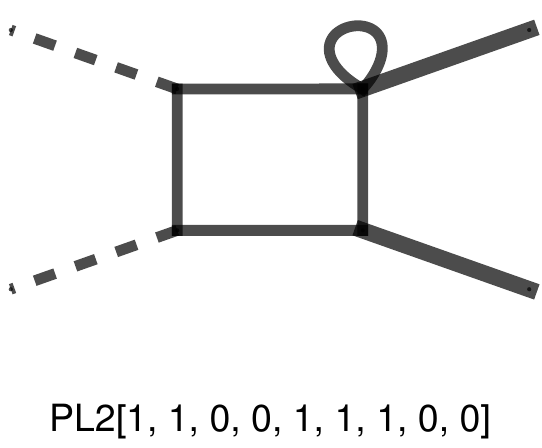} 
	\end{minipage}
	\begin{minipage}{0.16\textwidth} 
		\centering 
		\includegraphics[width=\textwidth]{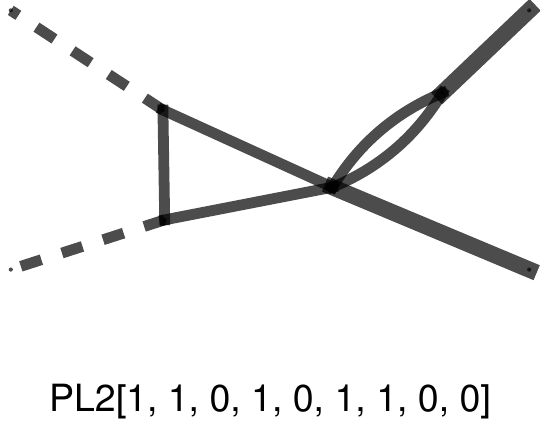} 
	\end{minipage}
	\begin{minipage}{0.16\textwidth} 
		\centering 
		\includegraphics[width=\textwidth]{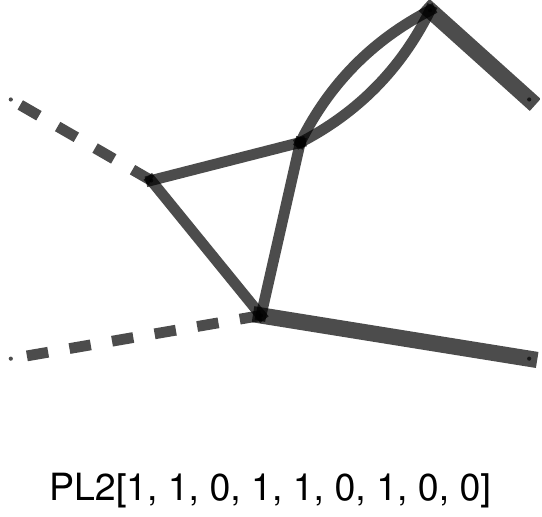} 
	\end{minipage}

    \begin{minipage}{0.16\textwidth} 
		\centering 
		\includegraphics[width=\textwidth]{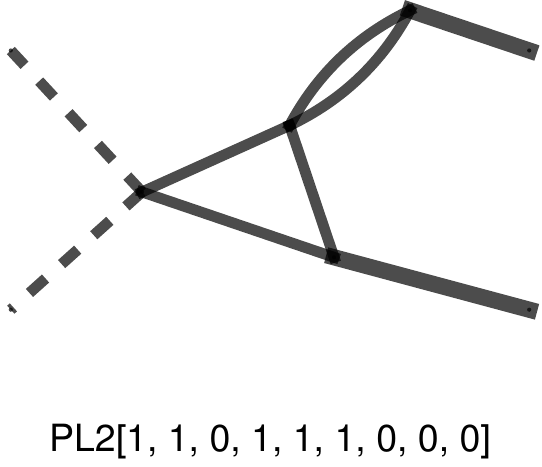} 
	\end{minipage}
    \begin{minipage}{0.16\textwidth} 
		\centering 
		\includegraphics[width=\textwidth]{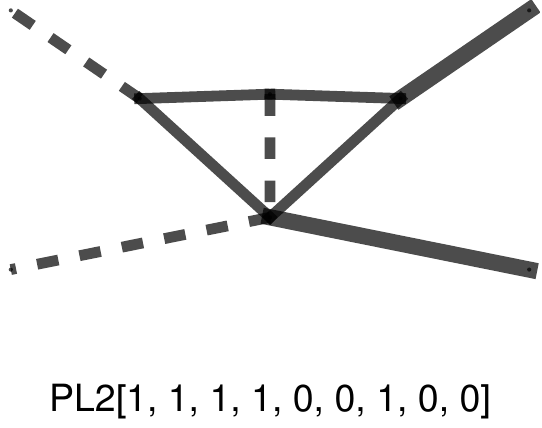} 
	\end{minipage}
    \begin{minipage}{0.16\textwidth} 
		\centering 
		\includegraphics[width=\textwidth]{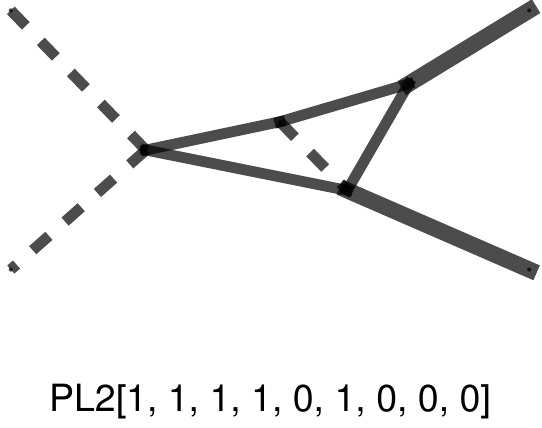} 
	\end{minipage}
	\begin{minipage}{0.16\textwidth} 
		\centering 
		\includegraphics[width=\textwidth]{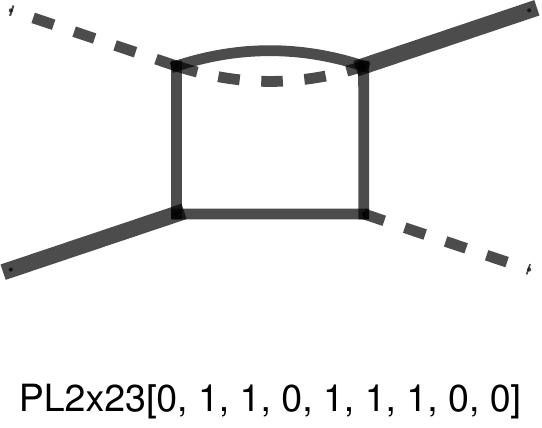} 
	\end{minipage}
	\begin{minipage}{0.16\textwidth} 
		\centering 
		\includegraphics[width=\textwidth]{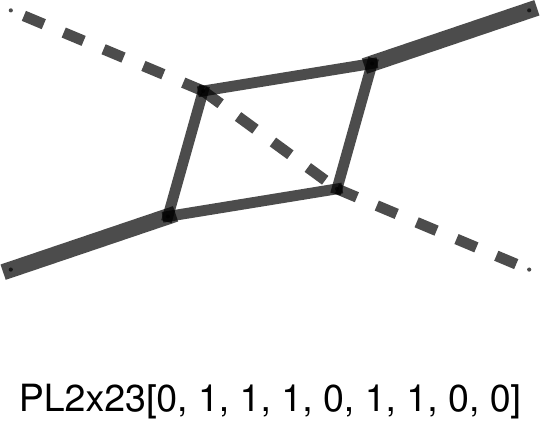} 
	\end{minipage}
	\begin{minipage}{0.16\textwidth} 
		\centering 
		\includegraphics[width=\textwidth]{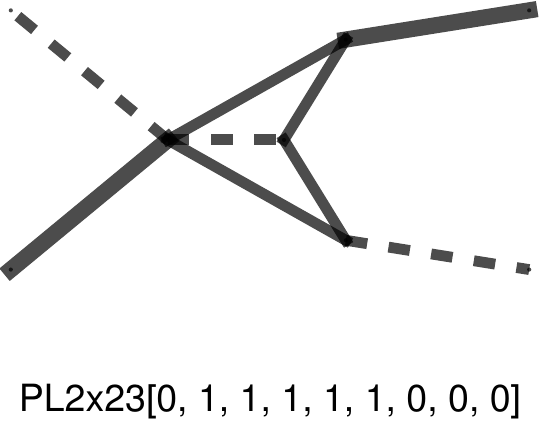} 
	\end{minipage}
    
    \begin{minipage}{0.16\textwidth} 
		\centering 
		\includegraphics[width=\textwidth]{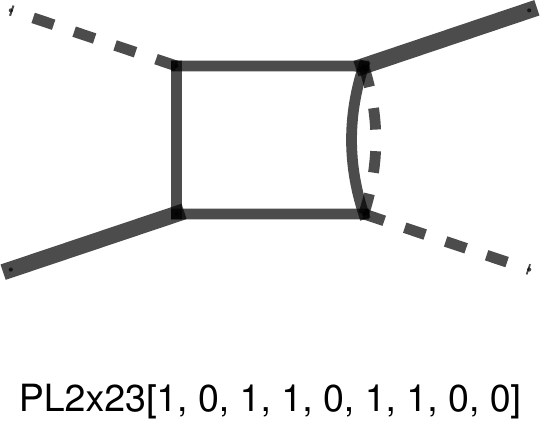} 
	\end{minipage}
    \begin{minipage}{0.16\textwidth} 
		\centering 
		\includegraphics[width=\textwidth]{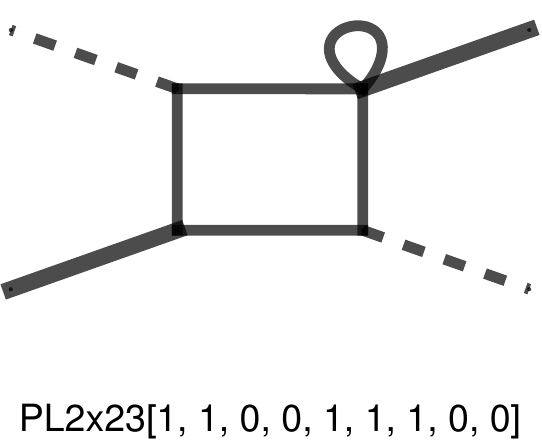} 
	\end{minipage}
     \begin{minipage}{0.16\textwidth} 
		\centering 
		\includegraphics[width=\textwidth]{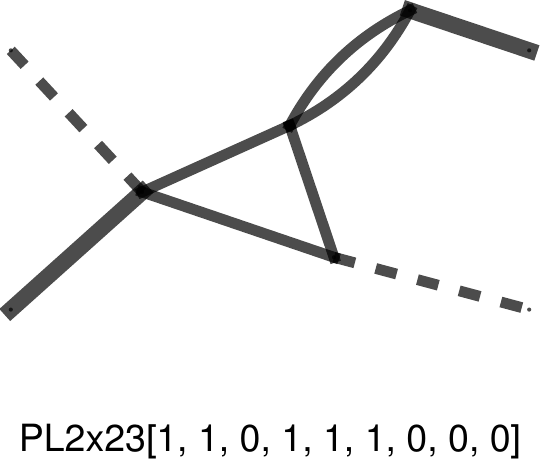} 
	\end{minipage}
	\begin{minipage}{0.16\textwidth} 
		\centering 
		\includegraphics[width=\textwidth]{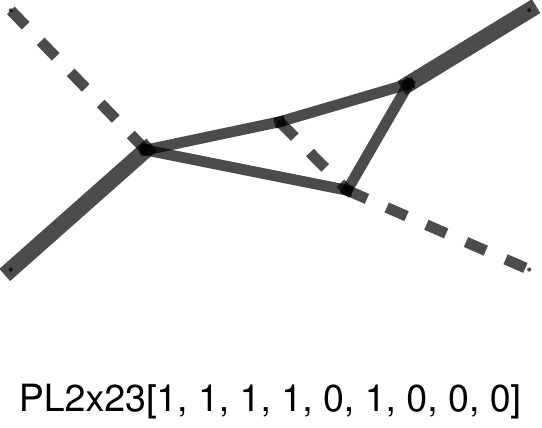} 
	\end{minipage}
	\begin{minipage}{0.16\textwidth} 
		\centering 
		\includegraphics[width=\textwidth]{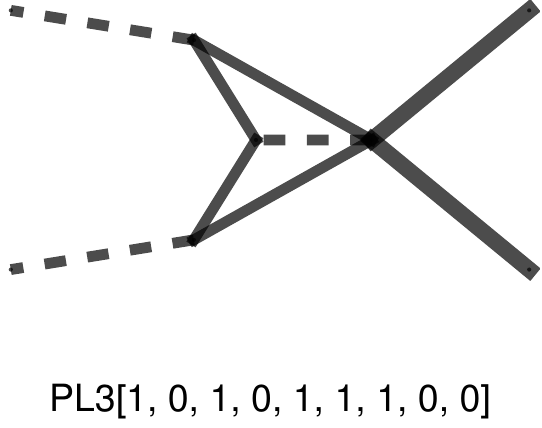} 
	\end{minipage}
	\begin{minipage}{0.16\textwidth} 
		\centering 
		\includegraphics[width=\textwidth]{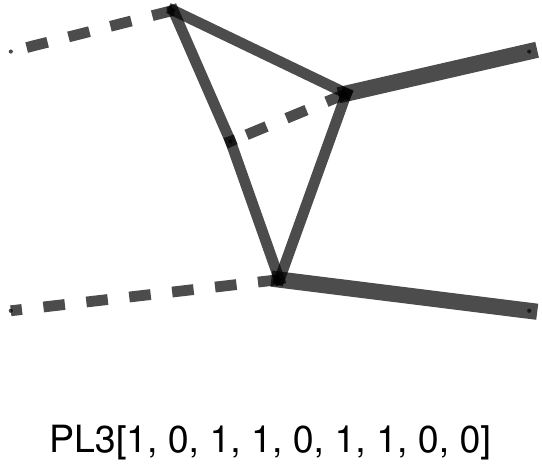} 
	\end{minipage}
\end{figure}

\begin{figure}[H]\ContinuedFloat
    \begin{minipage}{0.16\textwidth} 
		\centering 
		\includegraphics[width=\textwidth]{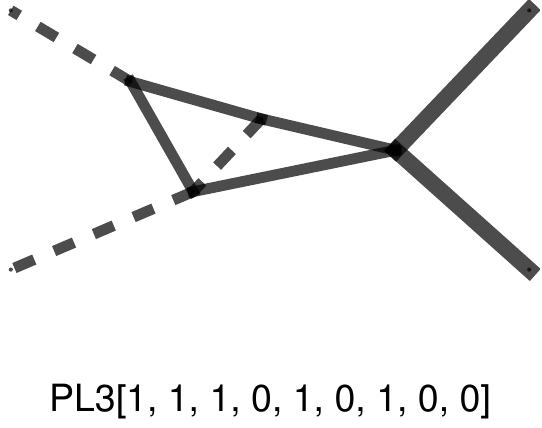} 
	\end{minipage}
    \begin{minipage}{0.16\textwidth} 
		\centering 
		\includegraphics[width=\textwidth]{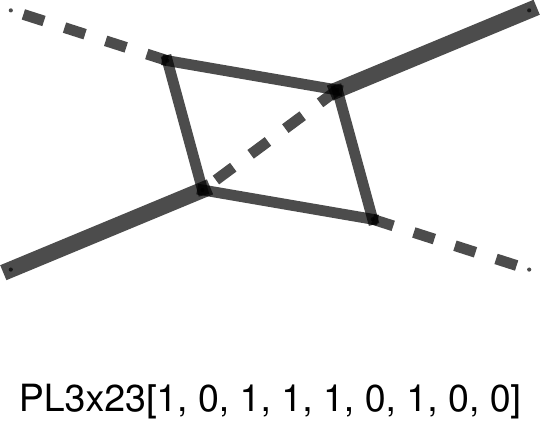} 
	\end{minipage}
    \begin{minipage}{0.16\textwidth} 
		\centering 
		\includegraphics[width=\textwidth]{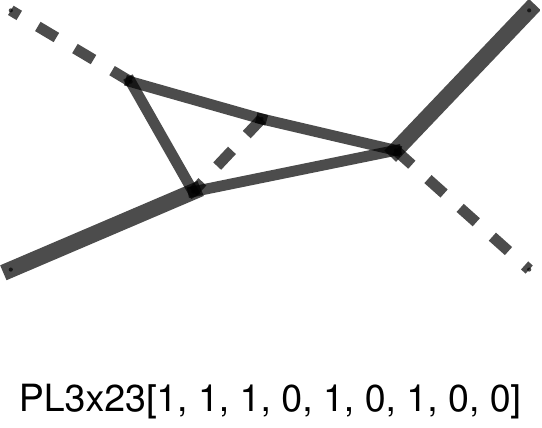} 
	\end{minipage}
	\begin{minipage}{0.16\textwidth} 
		\centering 
		\includegraphics[width=\textwidth]{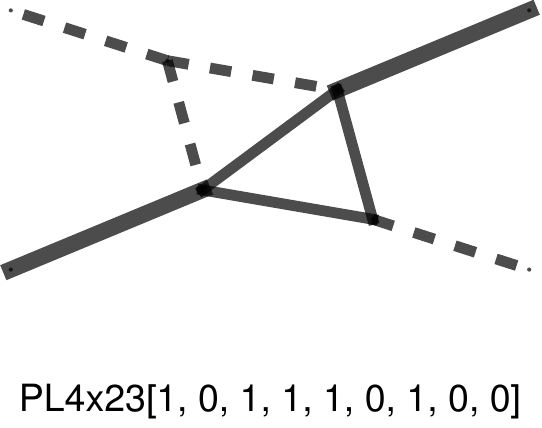} 
	\end{minipage}
    \begin{minipage}{0.16\textwidth} 
		\centering 
		\includegraphics[width=\textwidth]{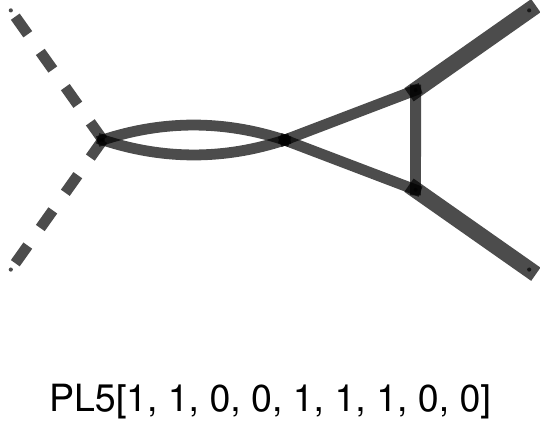} 
	\end{minipage}
	\begin{minipage}{0.16\textwidth} 
		\centering 
		\includegraphics[width=\textwidth]{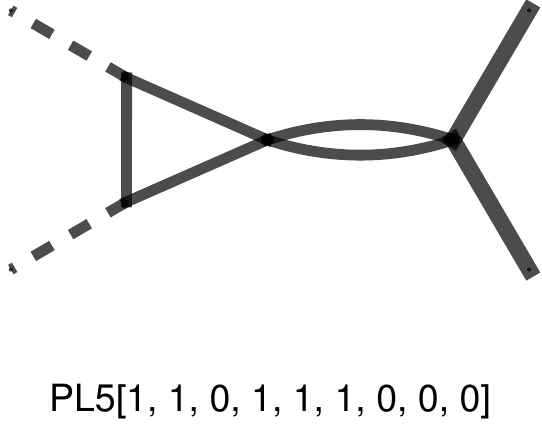} 
	\end{minipage}
   
    \begin{minipage}{0.16\textwidth} 
		\centering 
		\includegraphics[width=\textwidth]{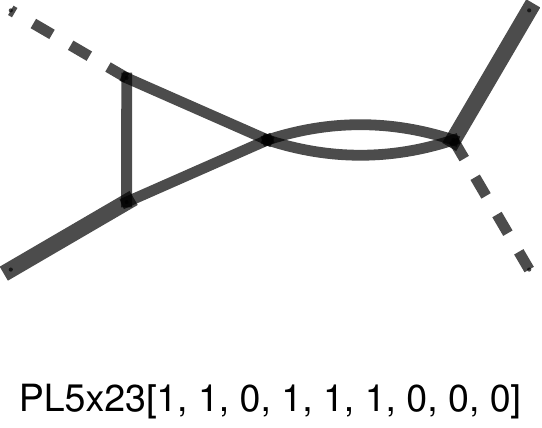} 
	\end{minipage}
    \begin{minipage}{0.16\textwidth} 
		\centering 
		\includegraphics[width=\textwidth]{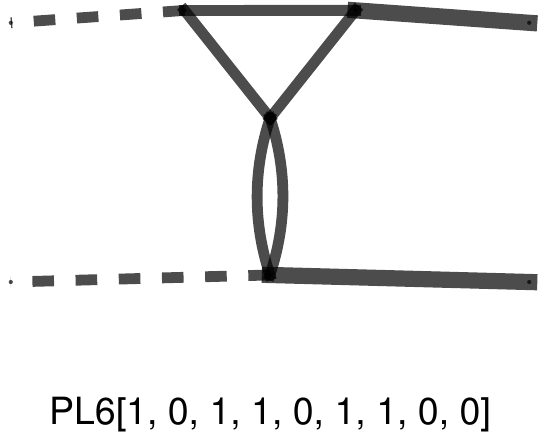} 
	\end{minipage}
    \begin{minipage}{0.16\textwidth} 
		\centering 
		\includegraphics[width=\textwidth]{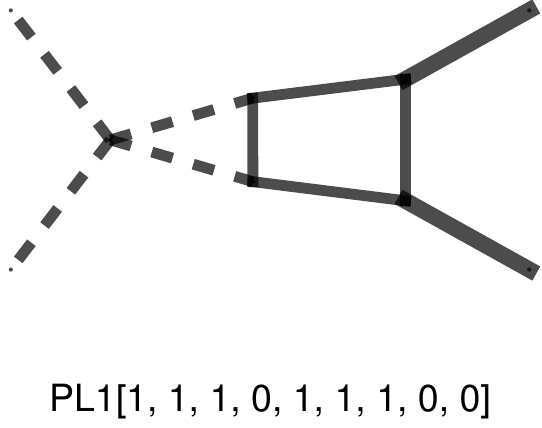} 
	\end{minipage}
	\begin{minipage}{0.16\textwidth} 
		\centering 
		\includegraphics[width=\textwidth]{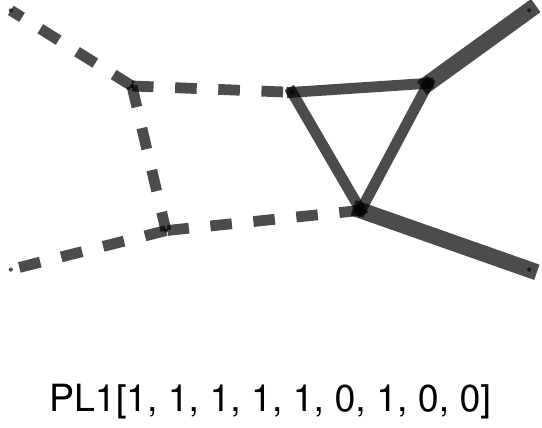} 
	\end{minipage}
	\begin{minipage}{0.16\textwidth} 
		\centering 
		\includegraphics[width=\textwidth]{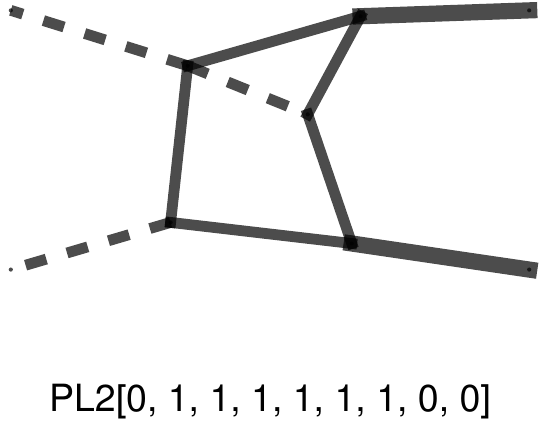} 
	\end{minipage}
	\begin{minipage}{0.16\textwidth} 
		\centering 
		\includegraphics[width=\textwidth]{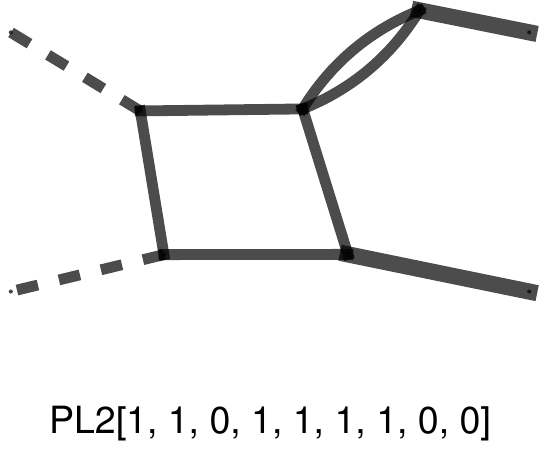} 
	\end{minipage}

    \begin{minipage}{0.16\textwidth} 
		\centering 
		\includegraphics[width=\textwidth]{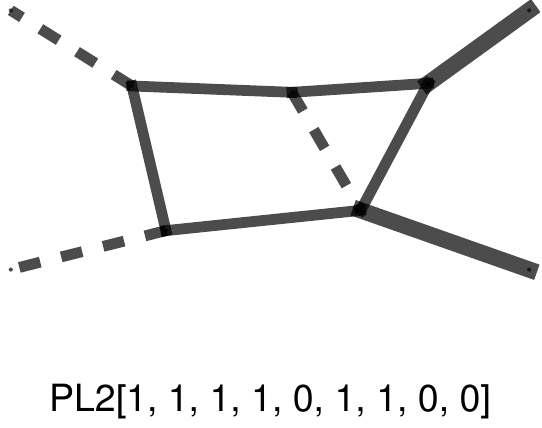} 
	\end{minipage}
    \begin{minipage}{0.16\textwidth} 
		\centering 
		\includegraphics[width=\textwidth]{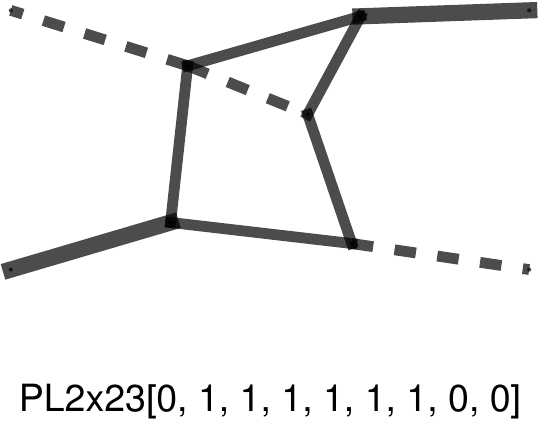} 
	\end{minipage}
    \begin{minipage}{0.16\textwidth} 
		\centering 
		\includegraphics[width=\textwidth]{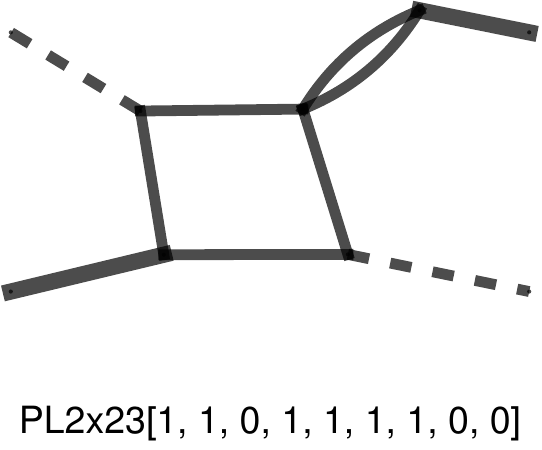} 
	\end{minipage}
	\begin{minipage}{0.16\textwidth} 
		\centering 
		\includegraphics[width=\textwidth]{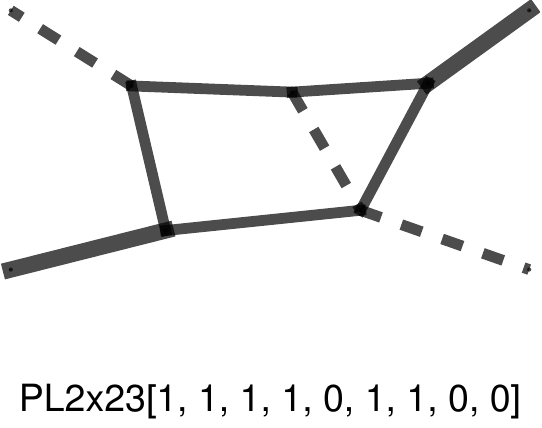} 
	\end{minipage}
	\begin{minipage}{0.16\textwidth} 
		\centering 
		\includegraphics[width=\textwidth]{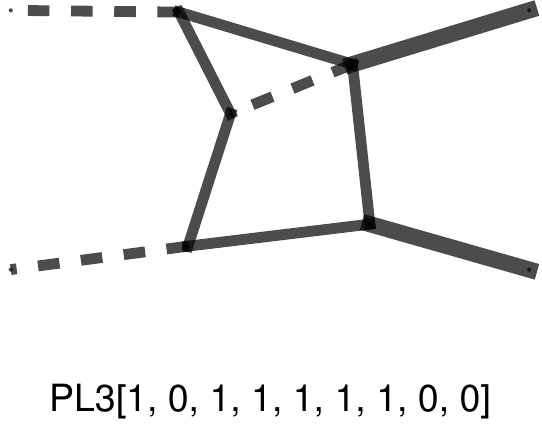} 
	\end{minipage}
	\begin{minipage}{0.16\textwidth} 
		\centering 
		\includegraphics[width=\textwidth]{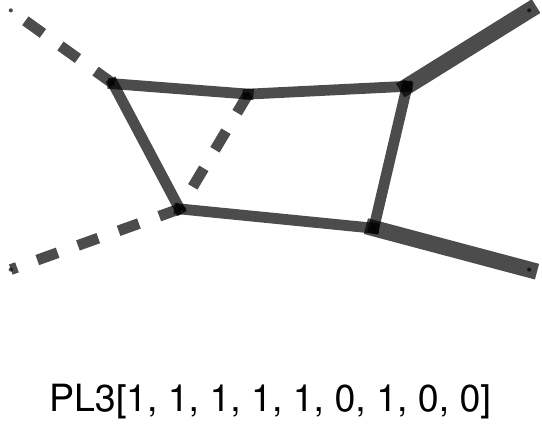} 
	\end{minipage}
    
    \begin{minipage}{0.16\textwidth} 
		\centering 
		\includegraphics[width=\textwidth]{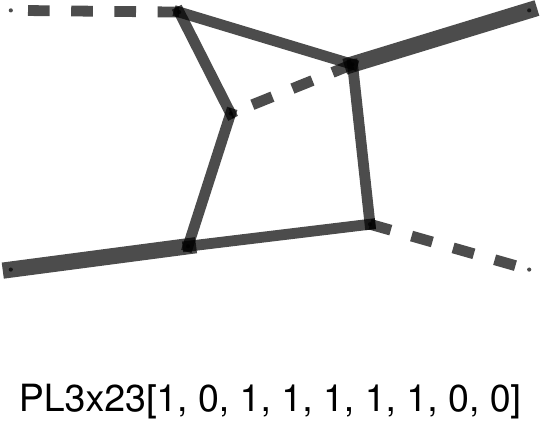} 
	\end{minipage}
    \begin{minipage}{0.16\textwidth} 
		\centering 
		\includegraphics[width=\textwidth]{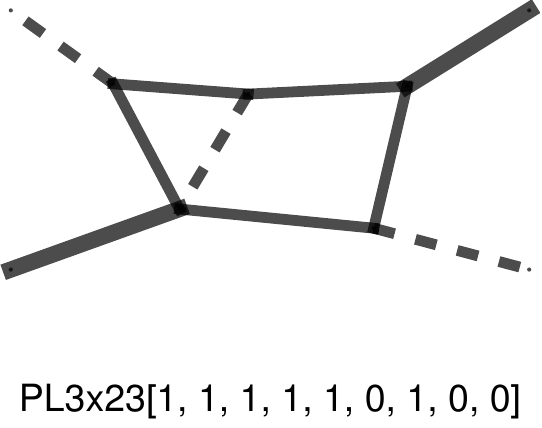} 
	\end{minipage}
    \begin{minipage}{0.16\textwidth} 
		\centering 
		\includegraphics[width=\textwidth]{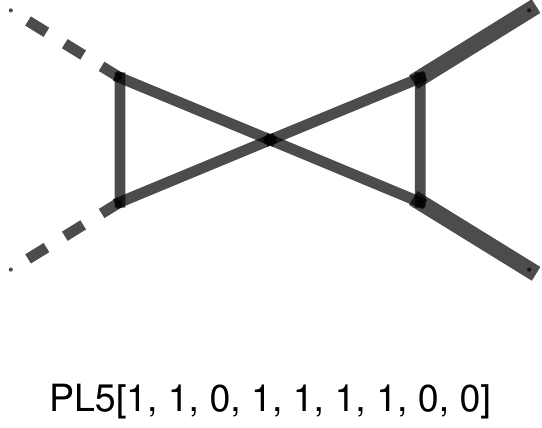} 
	\end{minipage}
	\begin{minipage}{0.16\textwidth} 
		\centering 
		\includegraphics[width=\textwidth]{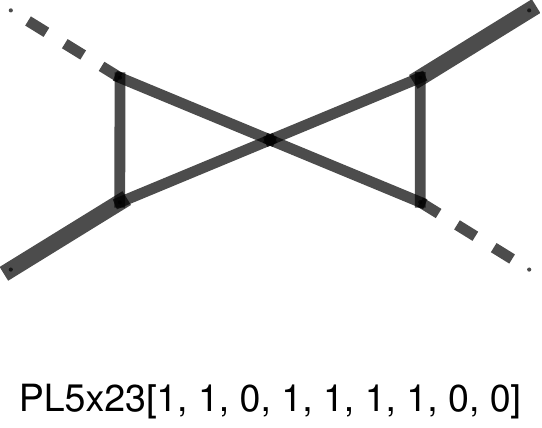} 
	\end{minipage}
	\begin{minipage}{0.16\textwidth} 
		\centering 
		\includegraphics[width=\textwidth]{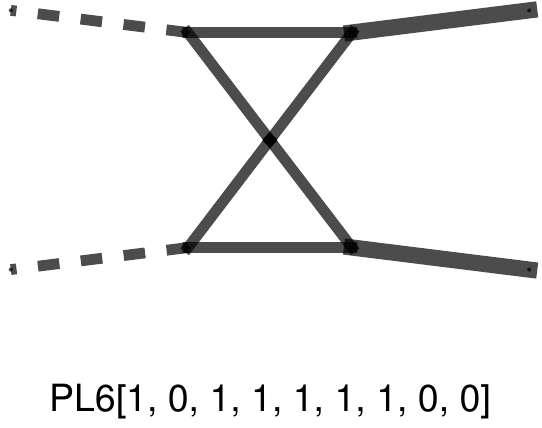} 
	\end{minipage}
	\begin{minipage}{0.16\textwidth} 
		\centering 
		\includegraphics[width=\textwidth]{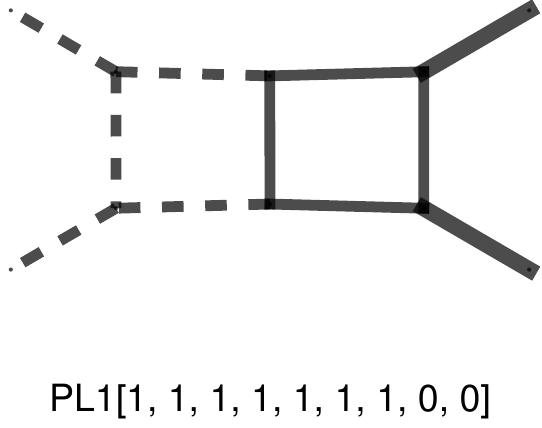} 
	\end{minipage}
    
    \begin{minipage}{0.16\textwidth} 
		\centering 
		\includegraphics[width=\textwidth]{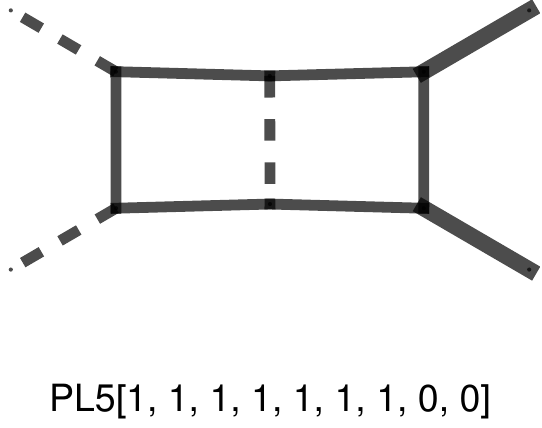} 
	\end{minipage}
    \begin{minipage}{0.16\textwidth} 
		\centering 
		\includegraphics[width=\textwidth]{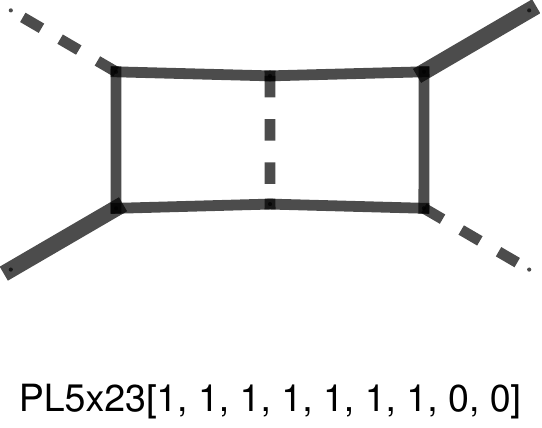} 
	\end{minipage}
    \begin{minipage}{0.16\textwidth} 
		\centering 
		\includegraphics[width=\textwidth]{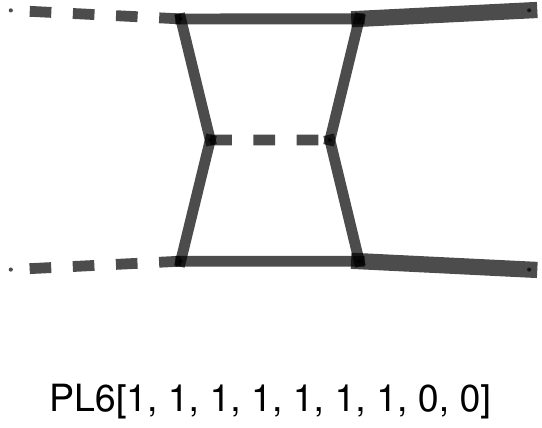} 
	\end{minipage}
    \caption{Two-loop planar master integrals without dots and scalar products are displayed here. The external solid lines and dashed lines represent the massive Higgs and massless gluons, respectively. With the same convention the internal solid lines and dashed lines represent the massive and massless propagators.}
    \label{Fig:MIs}
\end{figure}

\section{Multiple polylogarithms}
\label{appendixB}

Multiple polylogarithms (MPLs) are generalizations of harmonic polylogarithms(HPLs), Nielsen polylogarithms and the classical polylogarithms to multiple scales. They satisfy two different algebra structures, shuffle algebra and stuffle algebra. Here, we will provide the definition of MPLs and some properties that will be utilized in this work. More detailed information can be found in Refs.~\cite{Vollinga:2004sn,Duhr:2019tlz}.

The MPLs are defined recursively through the iterative integral
\begin{equation}
    G(a_1,...,a_n;z)=\,\int_0^z\,\frac{d t}{t-a_1}\,G(a_2,...,a_n;t)\,,\\
    \label{eq:def mpl}
\end{equation}
where $a_i$ and $z$ are complex variables. The vector $\vec a = (a_1,...,a_n)$ is called the weight vector, and the length $n$ is called the weight. In special case that the weight is 0 or all the $a_i$ are zero, they have
\begin{equation}
   G(;z) = 1  \,, \quad G(\underbrace{0,...,0}_{n\,\text{times}};z) = \frac{1}{n!}\,\log^n z\,.
\end{equation}
In the case that $a_n \neq 0$, according to scaling relation, they satisfy
\begin{equation}
    G(a_1,...,a_n;z) \, = \, G(x\,a_1,..,x\,a_n;x\,z) \,
    \label{eq:mpl rule1}
\end{equation}
with $x \neq 0$.

MPLs obey the shuffle algebra, which allows the product of two MPLs sharing the same variable $z$ with weight $w_1$ and $w_2$ to be expressed as a sum of MPLs with an overall weight of $w_1+w_2$ 
\begin{equation}
    G(\vec{a_1};z) \,  G(\vec{a_2};z) \, = \, G(\vec{a_1} \shuffle \vec{a_2};z)\,,
    \label{eq:mpl shuffle}
\end{equation}
where $\vec{a_1} \shuffle \vec{a_2}$ is their shuffle product, which is the sum of all possible permutations of the elements in $\vec{a_1}$ and $\vec{a_2}$ arranged in a way that does not change the internal order of them. For example,
\begin{eqnarray}
    G(a,b;z) \, G(c,d;z) &=& G(a,b,c,d;z) + G(a,c,b,d;z) + G(a,c,d,b;z) \nonumber\\
      &+& G(c,a,b,d;z) + G(c,a,d,b;z) + G(c,d,a,b;z)\,.
\end{eqnarray}
One way to prove Eq.~(\ref{eq:mpl shuffle}) is by the relation
\begin{equation}
    \int\limits_{0}^{z}d x_1\int\limits_{0}^{z}d x_2=\int\limits_{0}^{z}d x_1\int\limits_{0}^{x_1}d x_2+\int\limits_{0}^{z}d x_2\int\limits_{0}^{x_2}d x_1\,.
\end{equation}
Using the shuffle relations, one can obtain
\begin{eqnarray}
    G(a_1,...,a_n,\underbrace{0,...,0}_{k-j};z) &=& \frac{1}{k-j} G(0;z) \, G(a_1,...,a_j,\underbrace{0,...,0}_{k-j-1};z)   \nonumber \\
     &-& \frac{1}{k-j} \sum_{(s_1,...,s_j)} G(s_1,...,s_j,a_j,\underbrace{0,...,0}_{k-j-1};z) \,, 
     \label{eq:mpl rule2}
\end{eqnarray}
where $(s_1,...,s_j)=(a_1,...,a_{j-1})\shuffle (0)$. By iteratively applying the above equation, one can always express an MPL as a combination of MPLs where all elements in $\vec{a}$ are 0 or the last element in $\vec{a}$ is non-zero. In Sec~\ref{sec:Bc}, we use $\tt{GiNaC}$ library to compute the numerical values of MPLs. We express all MPLs into the form $G(a'_1,...,a'_n;1)$ by applying Eq.~(\ref{eq:mpl rule1}) and Eq.~(\ref{eq:mpl rule2}) to facilitate the definition of the branch cut.

The integration and differentiation of MPLs have been automated in the package $\tt{PolyLogTools}$, and further details can be found in its documentation.


 \bibliographystyle{JHEP}
 \bibliography{biblio.bib}


\end{document}